\documentclass[12pt,preprint]{aastex}

\shorttitle{X-ray Background with {\em Swift}/BAT}
\shortauthors{Ajello et al.}

\begin{document}

\title{Cosmic X-ray background and Earth albedo Spectra with {\em Swift}/BAT}

\author{M. Ajello\altaffilmark{1},
 J. Greiner\altaffilmark{1}, G. Sato\altaffilmark{2},
D. R. Willis\altaffilmark{1},
G. Kanbach\altaffilmark{1},  A. W. Strong\altaffilmark{1},
R. Diehl\altaffilmark{1}, G. Hasinger\altaffilmark{1}, 
N. Gehrels\altaffilmark{2}, 
C. B. Markwardt\altaffilmark{2} 
and J. Tueller\altaffilmark{2}
}
%%%%%%%%%%%%%%%%%%%%%%%%%%%%%%%%%%%%%%%%%%%%%%%%%%%%%%%%%%%%%%%%%%%%%%%%%%%%
\altaffiltext{1}{Max-Planck f\"ur Extraterrestrische Physik, Postfach 1312, 
85741, Garching, Germany}
\altaffiltext{2}{Astrophysics Science Division, Mail Code 661, NASA Goddard Space Flight Center, Greenbelt, MD 20771, USA }
\email{majello@mpe.mpg.de}

\begin{abstract}

We use {\em Swift}/BAT Earth occultation data 
at different geomagnetic latitudes to derive 
a sensitive measurement of the Cosmic X-ray background (CXB) 
and of the Earth albedo emission in the 15--200\,keV band.
We compare our CXB spectrum with recent ({\it INTEGRAL}, BeppoSAX) and
past results (HEAO-1) and find good agreement. Using an independent
measurement of the CXB spectrum we are able to confirm our results.
This study shows that the BAT CXB spectrum has a normalization 
$\sim8\pm3$\,\%
larger than the HEAO-1 measurement.
The BAT accurate Earth albedo spectrum can be used to predict 
the level of photon background for satellites 
in low Earth and mid inclination orbits.
\end{abstract}



\section{Introduction}
There is a general consensus that the cosmic X-ray
background (CXB), discovered more than 40 years ago \citep{giacconi62},
is produced by integrated emission of extra-galactic point sources.
The deepest X-ray surveys to date \citep{giacconi02,alexander03,hasinger04}
have shown that up to virtually 100\% of the $<$ 2\,keV 
CXB radiation  is accounted for by Active Galactic Nuclei (AGN) 
hosting accreting super-massive black holes (SMBHs).
However, the fraction of CXB emission resolved into AGNs declines
with energy   being $<50$\% above 6\,keV \citep{worsley05}.
The unresolved  component  may be attributed to
the emission of a yet undetected population of highly-absorbed  AGN.
These AGN should be characterized by having column densities 
$\sim10^{24}$ H-atoms cm$^{-2}$ and a space density peaking at redshift 
below 1 \citep{worsley05}.
Such a population of Compton-thick AGN is invoked by population
synthesis models \citep[e.g.][]{comastri95,treister05,gilli07}
to reproduce the peak of the CXB emission at 30\,keV \citep{marshall80}.

Thus, an accurate measurement of the CXB spectrum in the 15--200\,keV
energy range is important to assess and constrain
the number density of Compton-thick AGNs. Such measurements are 
complicated by the fact that instruments sensitive in this energy 
range are dominated by internal detector background and are
not designed to measure the CXB spectrum directly (excluding HEAO-1 A2).
The typical approach is to produce an $ON-OFF$ measurement, where
taking the difference between the $ON$ and the $OFF$ 
pointings eliminates the internal background component.

There are different methods to obtain a suitable $OFF$ observation;
the HEAO-1 measurement of the CXB spectrum in the 13--180\,keV range
was obtained by blocking the aperture with a movable CsI  crystal.
Also the Earth disk can be used to modulate the CXB emission.
This approach is  the one used in recent CXB intensity measurements
performed by {\it INTEGRAL} and BeppoSAX \citep{churazov07,frontera07}.

Here we report on two independent measurements  of the CXB emission
using  {\em Swift}/BAT. For the first method, we use the Earth 
occultation technique similarly to the {\it INTEGRAL} and BeppoSAX analyses while
for the second one we make use of the spatial distribution of the BAT 
background.

The structure of the paper is as follows. In $\S$~\ref{sec:obs}
we present the details of the observations and describe the BAT background
components. We also derive a rate-rigidity
relation which is fundamental for suppressing the background variability
due to Cosmic Rays (CRs).
In $\S$~\ref{sec:analysis} we present the details of the Earth's
occultation episodes undergone by BAT and the analysis method for
the occultation measurement. In $\S$~\ref{sec:err}, we discuss
all sources of uncertainties which affect our occultation measurement
which is then presented in $\S$~\ref{sec:results}. 
The alternative measurement used to verify the results of the occultation 
analysis is reported in $\S$~\ref{sec:quad}.
We discuss the broad band properties of the CXB and Earth spectra
in $\S$~\ref{sec:disc}.
Finally, the last section summarizes our findings.

%%%%%%%%%%%%%%%%%%%%%%%%%%%%%%%%%%%%%%%%%%%%%%%%%%%%%%%%%%%%%%%%%%%%%%
%       Observations
%
\section{Observations}\label{sec:obs}
The Burst Alert Telescope \citep[BAT;][]{barthelmy05}, aboard the 
{\it Swift} mission \citep{gehrels04},  launched by NASA in 2004,
 represents
a major improvement in sensitivity for X-ray imaging of the hard X-ray sky.
BAT is a coded mask telescope with a wide field of view 
(FOV, $120^{\circ} \times 90^{\circ}$ partially coded) 
sensitive in the hard X-ray domain (15-200\,keV).
BAT's  main purpose is to locate and to study Gamma-Ray Bursts (GRBs).
While  chasing new GRBs, 
BAT surveys the hard X-ray sky  with an unprecedented sensitivity.
Thanks to its wide FOV  and its  pointing strategy, 
BAT monitors continuously a large fraction of the sky (up to 80\%) every day.

The {\it Swift} satellite constraints require that
the pointing direction be at least 30$^{\circ}$ above
the Earth's horizon. Nevertheless, 
due to its extent, it may happen that the 
Earth disk occults a substantial portion (up to 30\%) of the BAT FOV.
Moreover, BAT survey data include episodes of large
occultation (up to $\sim$70\%) caused by the Earth when the spacecraft
was in 'safe' mode.

We use 8 months of BAT data which constitutes a  well characterized 
dataset of BAT survey data \citep[see][for details]{ajello08a,ajello07b}
to study the different components of the BAT background.
Our first aim is to derive the BAT background spectrum in the infinite-rigidity
approximation. We then use all occultation episodes, as described in 
$\S$ \ref{sec:analysis}, to derive
a measurement of the CXB and the Earth  atmosphere spectra.

%
%%%%%%%%%%%%%%%%%%%%%%%%%%%%%%%%%%%%%%%%%%%%%%%%%%%%%%%%%%%%%%%%%%%%%%%
%       The BAT background
%
\subsection{The BAT background}
The BAT background is highly complex and structured; 
it exhibits variability dependent on both orbital position and 
pointing direction. BAT employes a graded-Z fringe shield 
to suppress the in-orbit background. The fringe shield, located
around and below the BAT detector plane, reduces the isotropic cosmic 
diffuse flux and the anisotropic Earth albedo by $\sim$95\% 
\citep{barthelmy05}. The two main background components 
are the CXB emission and the cosmic ray induced (prompt and delayed)
backgrounds. 

The CXB spectrum in the 3--400\,keV range is derived from HEAO-1 data. 
The following analytical approximation was suggested by \cite{gruber99}:
\begin{equation}
S_{CXB}(E) = \left\{
\begin{array}{lr}
7.877 E^{-0.29} e^{-E/41.13} &   3 < E < 60\,{\rm keV}  \\
0.0259 \left(\frac{E}{60}\right)^{-5.5} + 0.504\left( \frac{E}{60}\right)^{-1.58} +  0.0288 \left( \frac{E}{60}\right)^{-1.05} 
& E > 60\,{\rm keV} 
\end{array}
\right.
\label{eq:gruber}
\end{equation}
where  $S_{CXB}(E)$ is expressed in units of 
keV cm$^{-2}$ s$^{-1}$ sr$^{-1}$ keV$^{-1}$. Given the large FOV 
($\sim$1.4\,sr half coded), the CXB is the dominant background
component in BAT up to $\sim$50--60\,keV.

The prompt CR background is due to spallation effects
of incident CRs on the material of the spacecraft; since the Earth magnetic
field modulates the flux of incident CRs across the orbit,
such background component is expected to vary with the cut-off
rigidity R$_c$ (i.e., the minimum momentum an incident charged particle must 
have in order to penetrate into the Earth's magnetosphere).
The delayed component is caused by the  excitation of the materials from
the incident CR flux. This component  builds up 
on times short compared to the relevant decay lifetimes,
 then varies as the slower of the irradiation or the lifetime.
The Earth's magnetic field includes an indentation in the southern 
hemisphere 
called the South Atlantic Anomaly (SAA).
During each SAA passage, BAT experiences a sharp increase in count rate
due to the increase of the incident CR flux  and a delayed background
due to de-excitation of the spacecraft materials.

In order to discriminate the various components of the BAT background, 
we correlated the BAT whole array
rate (in each energy channel and normalized by the number of working detectors)
with   several orbital parameters.
The final goal is to derive a ``steady-state'' BAT background model which is 
unaffected by orbital variations.

%
%%%%%%%%%%%%%%%%%%%%%%%%%%%%%%%%%%%%%%%%%%%%%%%%%%%%%%%%%%%%%%%%%%%%%%%%%%%%
%                Data cuts
%
\subsubsection{Data Selections} \label{sec:cuts}
Our aim is to determine a rate-rigidity relation in order to extrapolate
the BAT array rates to the infinite-rigidity case; this allows us to
model the background variability  due  to the prompt and delayed
CR components.

First we selected the data  
excluding all observations where sources with signal-to-noise 
ratio (S/N) greater than 8 are detected. An 8$\sigma$ source produces
an increase in rate of less than 0.5\% in a typical 300\,s 
observation\footnotemark{}; thus
all point-like sources below this limit give a negligible contribution to
the background level.
\footnotetext{BAT survey observation have typically an exposure of 300\,s,
although shorter and longer exposures might exist.}

\begin{figure*}[ht!]
  \begin{center}
  \begin{tabular}{cc}
    \includegraphics[scale=0.8]{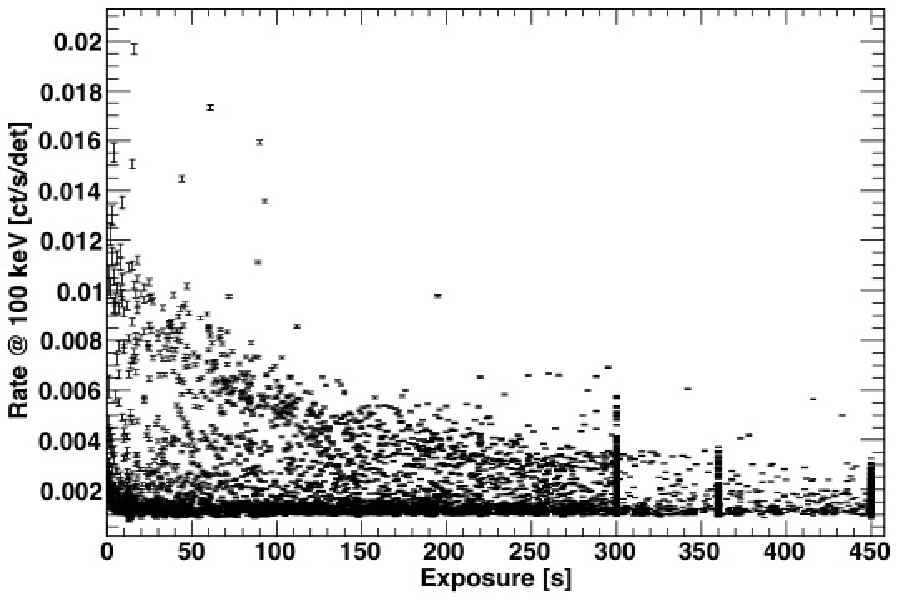} 
  	 \includegraphics[scale=0.8]{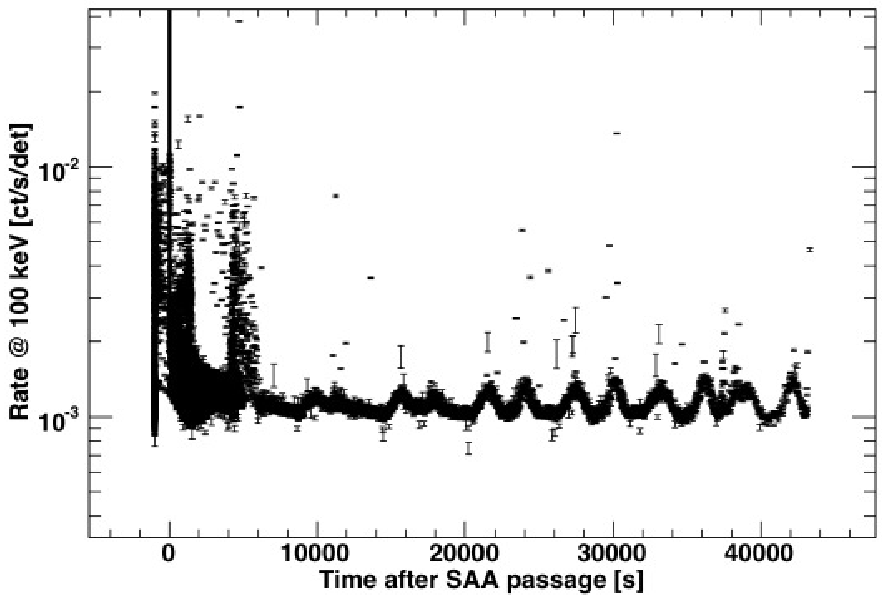}\\
\end{tabular}
  \end{center}
  \caption{
{\bf Left Panel:}
BAT rate at 100\,keV versus exposure time.  The ``truncated'' exposures below
300\,s are noisy. Despite the impression most ($\geq$90\,\%) of the observations
have exposure larger than (or equal)  300\,s.
{\bf Right Panel:}
BAT rate at 100\,keV versus time after each SAA passage. 
Note the sharp increase
in rate and decay behaviour when the spacecraft exits the SAA.
The second sharp peak at $\sim$4000\,s is due to a subsequent passage within
the tail of the SAA.
The rate modulation visible after $\sim$6000\,s is due to the
Earth's magnetosphere. Negative times are used for exposure taken within
the SAA.}
 \label{fig:saa}
\end{figure*}

The next step is to eliminate all observations whose exposure time is 
less than 300\,s. As Fig.~\ref{fig:saa} (left panel) shows, 
for exposures below 300\,s, 
the rates show a clear anti-correlation with exposure
time, with an increase of a factor $\sim$3 in 
the rates for few seconds of exposures. Exposures below  300\,s
are usually the result of a truncated observation
because: 1) BAT detects a GRB or 
2) BAT enters in the SAA and data acquisition is suspended.
For any of these reasons, data of truncated exposures are excluded by the 
present analysis because they are 
not representative of the average BAT background.

Whenever the spacecraft exits  the SAA, BAT experiences a rate
decline due to de-excitation of spacecraft materials.
As shown in Fig.~\ref{fig:saa} (right panel) the rates reach their normal
level after $\sim$5600\,s after each SAA passage. By excluding all
observations taken within $\sim$5600\,s of an SAA passage, we thus eliminate
short-lived radioactivity effects.

The BAT rates also show  a correlation with the angle between the Sun 
and the pointing direction. This correlation becomes visible
 at angles $>$120$^{\circ}$
and decreases with energy, disappearing at $70-80$\,keV.
The reason of this rate increase with the angle to the Sun 
is unclear. However, since the number of these
observations is small ($\sim$5\,\%), we decided to exclude them
from the present analysis.

The BAT effective area declines with energy and 
at $\sim$ 200 keV reduces to 1/5 of its peak value
at 50 keV; moreover, at these energies the fringe shield becomes partially
transparent. Thus, it is possible to use the high energy channels as a
``particle'' detector to monitor the background level of the instrument
(i.e. these channels do not yield much information on any celestial signal).
We found that imposing that the rate of the last energy channel (194\,keV--6.5\,MeV) be
in the range 10--20\,ct s$^{-1}$ eliminates roughly 1\% of the observations
which are outliers in all the correlations we have studied.

\begin{figure*}[ht!]
  \begin{center}
  \begin{tabular}{cc}
    \includegraphics[scale=0.43]{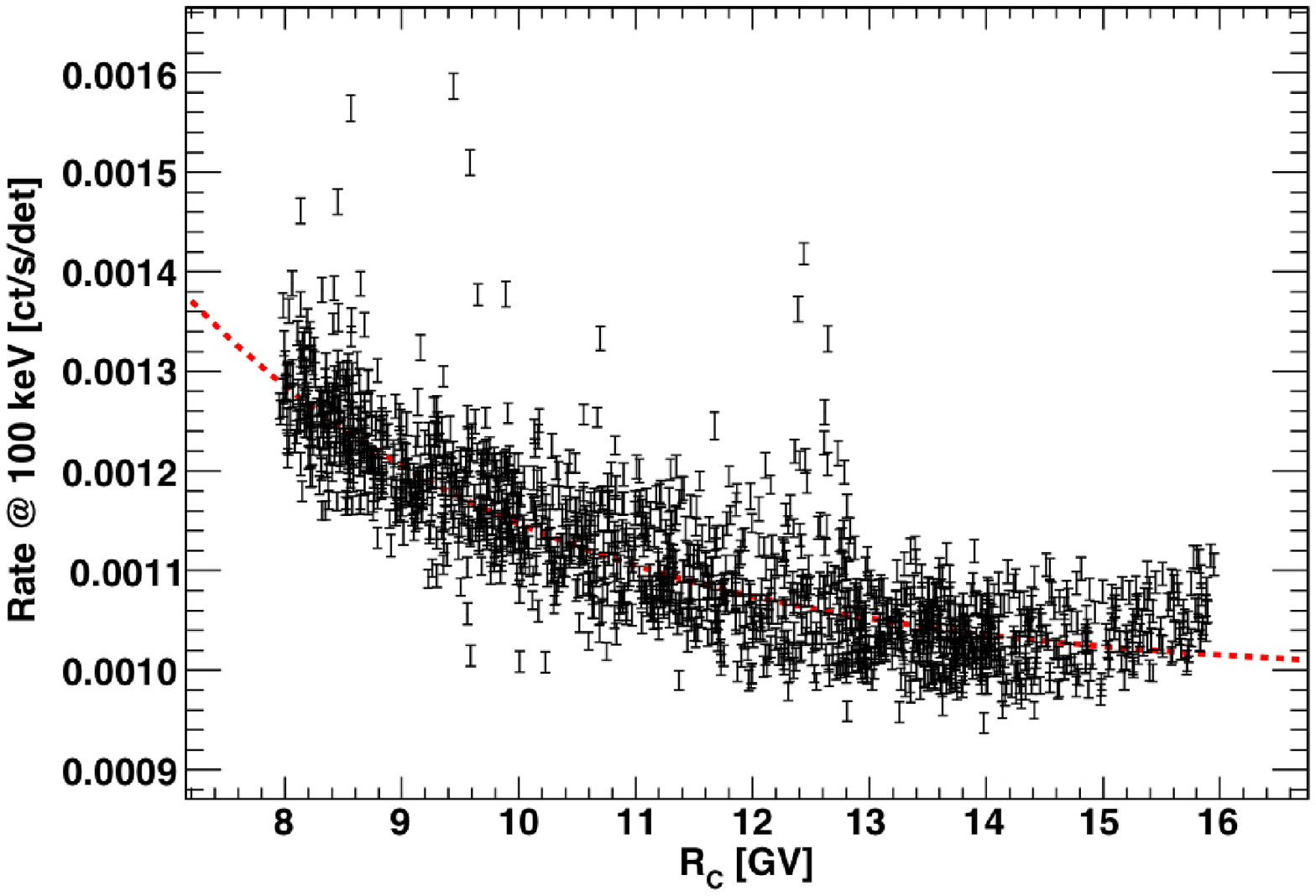} 
  	 \includegraphics[scale=0.43]{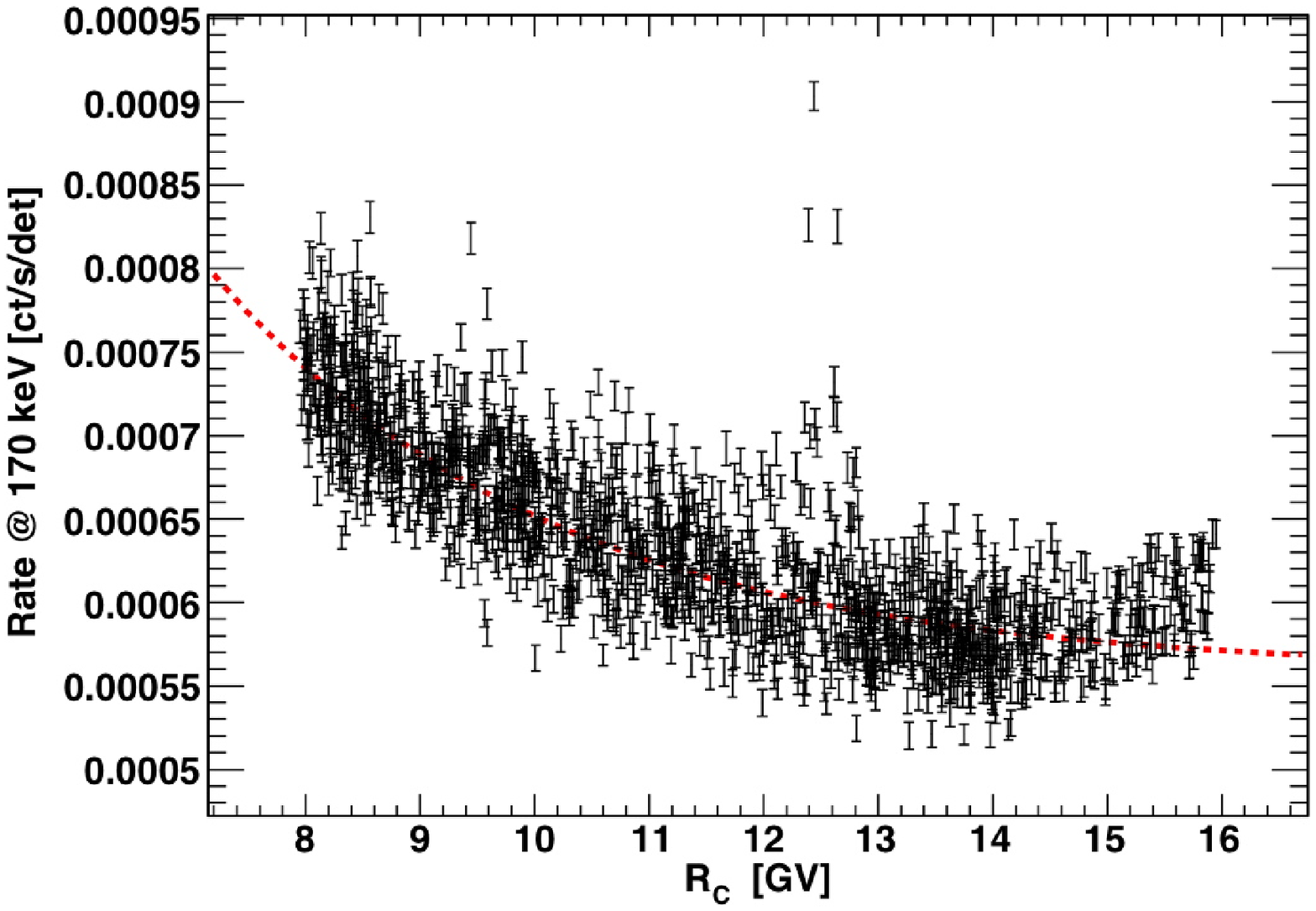}\\
\end{tabular}
  \end{center}
  \caption{
{\bf Left Panel:}
Correlation of BAT rate with the cut-off rigidity at 100\,keV. The dashed
line is the best fit using  an exponential plus a constant.
Only those  observations which passed the 
selection criteria explained in $\S$~\ref{sec:cuts} were used.
{\bf Right Panel:}
Correlation of BAT rate with the cut-off rigidity at 170\,keV. The dashed
line is a exponential plus a constant fit. The outliers present 
in both figures around 12\, GV are due to the spacecraft being in the 
vicinity  of the SAA. 
Only those  observations which passed the 
selection criteria explained in $\S$~\ref{sec:cuts} were used.
}
  \label{fig:rate_rig}
\end{figure*}

\begin{figure*}[ht!]
  \begin{center}
  \begin{tabular}{lc}
 \includegraphics[scale=0.43]{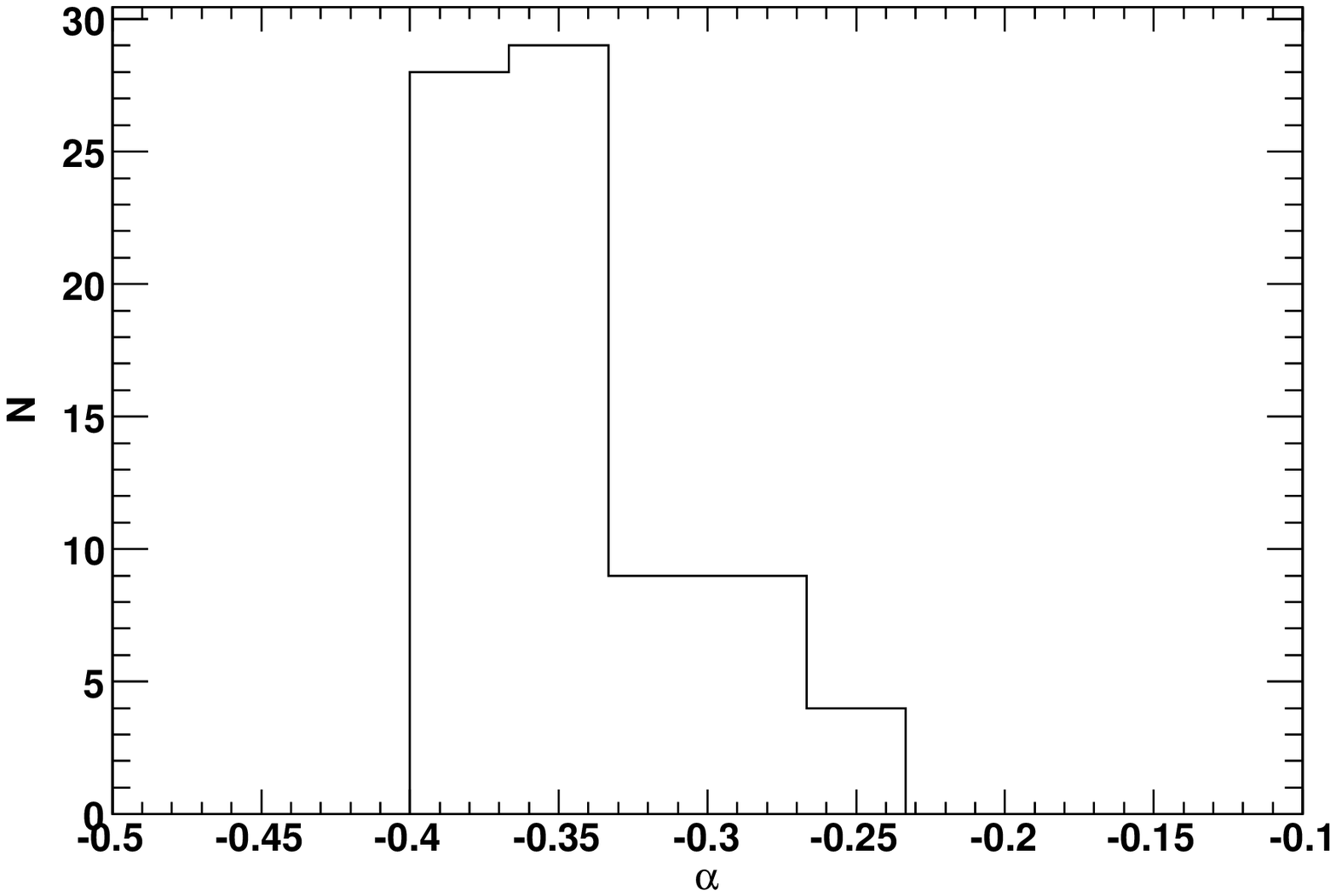} 
 \includegraphics[scale=0.43]{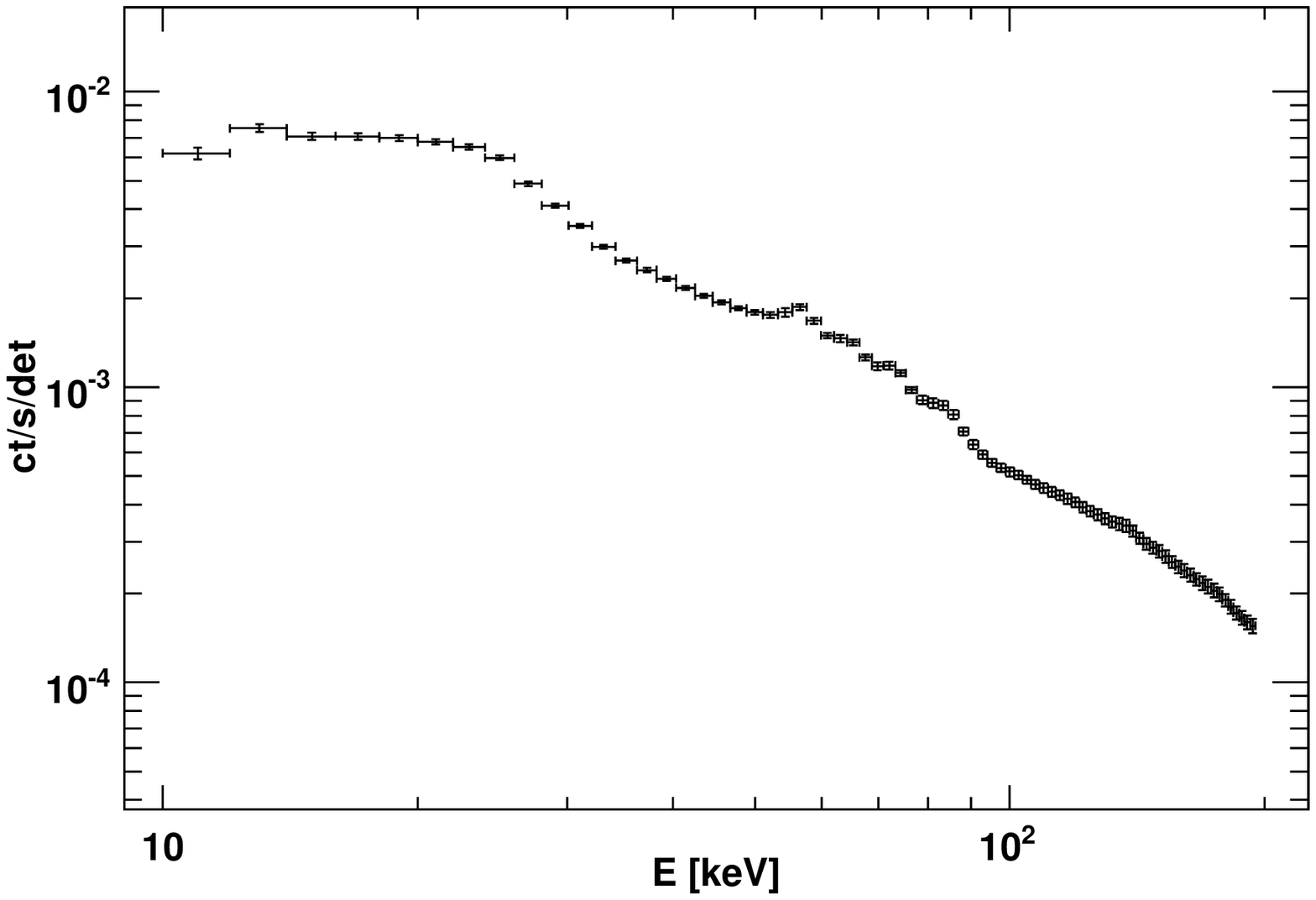} \\
\end{tabular}
  \end{center}
  \caption{
{\bf Left Panel:}
Distribution of the $\alpha$ values. The mean of -0.34 is in
good agreement with measurements from \cite{imhof76}.
{\bf Right Panel:}
BAT background spectrum extrapolated in the infinite rigidity case.
}
  \label{fig:err}
\end{figure*}

After these cuts, we find, as expected,
that the rates in each energy channel decrease
as a function of the cut-off rigidity R$_C$. This effect
is shown for two representative energy channels in Fig.~\ref{fig:rate_rig}.
We model this behavior with an exponential and a constant 
($Rate = C + Be^{\alpha R_C}$),
and fit this model to each 
energy channel.  The fitted constant $C$ provides an estimate
of the BAT rate in the infinite-rigidity extrapolation.
The 	distribution of the steepness of the rate increase with rigidity
($\alpha$ values), shown in left panel
of  Fig.~\ref{fig:err},
 has a mean of -0.34
and a RMS of 0.04, in perfect agreement  with previous measurements 
\citep{imhof76}. 

%%%%%%%%%%%%%%%%%%%%%%%%%%%%%% --------- Fig 2

%%%%%%%%%%%%%%%%%%%%%%%%%%%%%% --------- Fig 3

The BAT spectrum obtained by extrapolating the rates of each energy channel
to infinite rigidity is shown in the right panel of Fig.~\ref{fig:err}. 
The bumpiness between 60 and 100\,keV is due to the
numerous fluorescence emission lines from the fringe shield
\citep[see][for details]{willis02}.

%
%%%%%%%%%%%%%%%%%%%%%%%%%%%%%%%%%%%%%%%%%%%%%%%%%%%%%%%%%%%%%%%%%%%%%%%%%%
%
%                  Earth Occultation
%
\section{Earth occultation} \label{sec:analysis}
\begin{figure}[ht!]
  \begin{center}
  	 \includegraphics[scale=0.9]{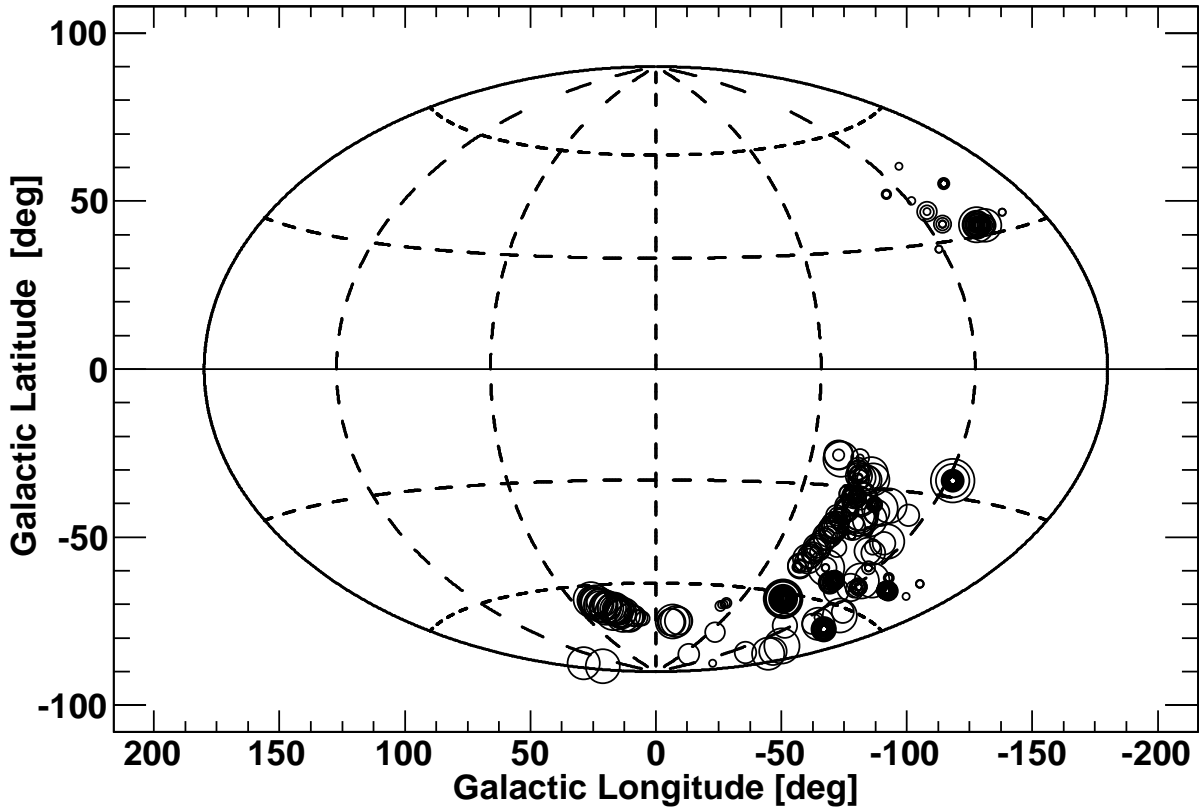}
  \end{center}
  \caption{Pointing directions for the occulted observations. Marker sizes
vary linearly with the occulted FOV fraction. Occultations vary
from 0.1\,\% to $\sim$80\,\%. In order to avoid contamination of the CXB signal
from the Galactic Ridge emission, we used only occultation episodes at
$|$b$|>$20$^{\circ}$ (data points already excised from this plot).}
\label{fig:occ_ptg}
\end{figure}

The {\it Swift} orbital constraints require that the  BAT pointing direction
be always at least 30$^{\circ}$ away from the Earth horizon.
This is  because the Earth is bright in Optical and X-rays, thus
it may damage the UV/Optical telescope (UVOT) and the X-ray telescope (XRT).
On May 31, June 12 and July 28, 2005, the {\em Swift} spacecraft entered into
'safe'-mode because of star tracker loss of lock. 
In safe-mode
operations, the XRT and UVOT telescopes are closed, but BAT still takes
data. The spacecraft remains in sun reference pointing 
until commanding
from the ground recovers {\em Swift} back to its normal status.
In the timespan between the safe-mode and the recovering operation,
the satellite uses the magnetometers and the sun sensor to derive
its pointing direction.
At least in the occasions mentioned
above\footnotemark{},
\footnotetext{A few episodes of Earth occultation were
found in BAT data,  but some did not pass the criteria 
explained in Sec. \ref{sec:cuts}}
the Earth passed through the BAT field of view (FOV).  Fig.~\ref{fig:occ_ptg}
shows the BAT pointing directions during the deep occultation episodes 
described here.

%%%%%%%%%%%%%%%%%%%%%%%%%%%%%% --------- Fig 4

\begin{figure*}[ht!]
  \begin{center}
  \begin{tabular}{cc}
    \includegraphics[scale=0.43]{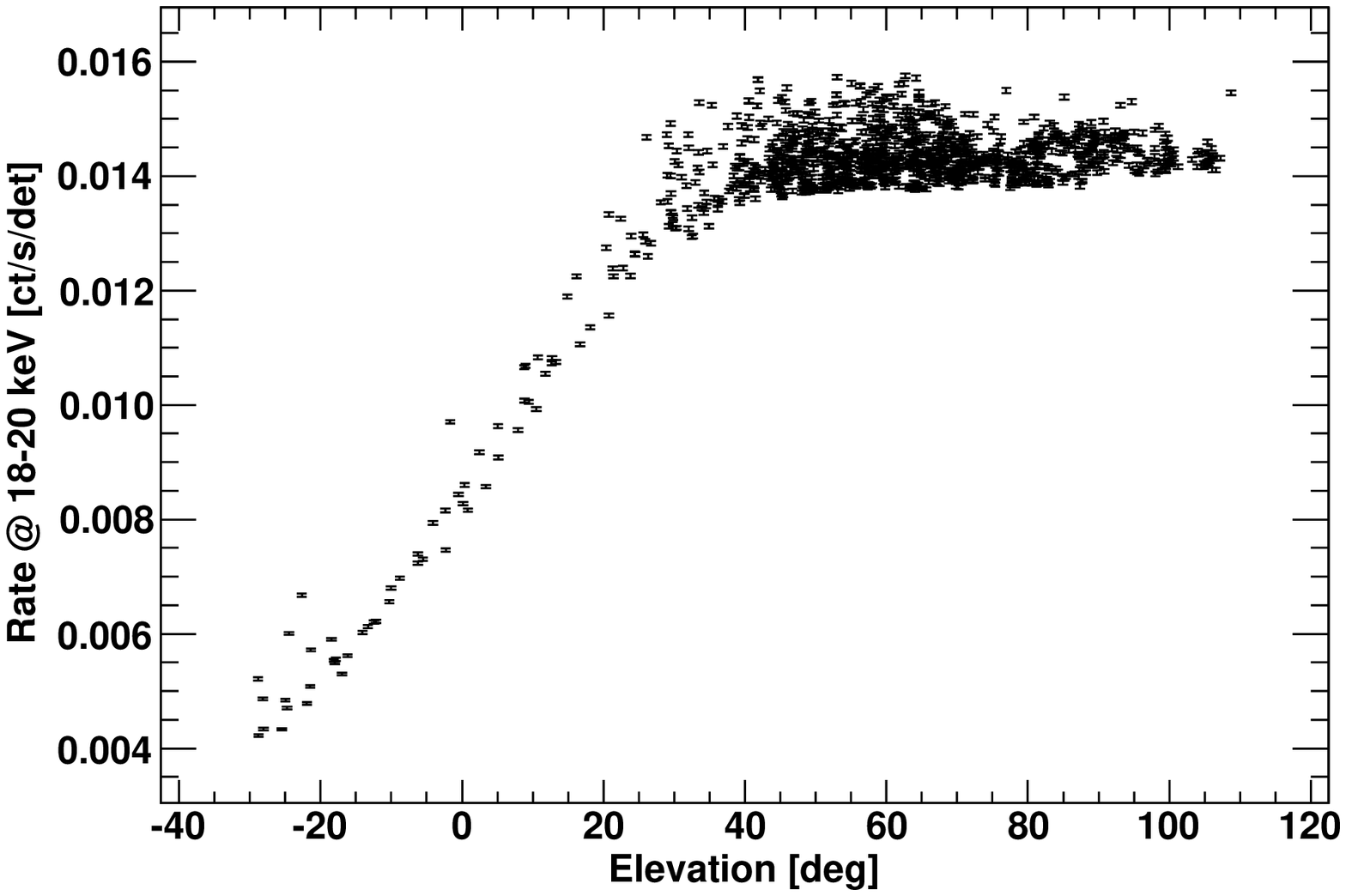} 
  	 \includegraphics[scale=0.40,trim=0 -38 0 0,clip=true]{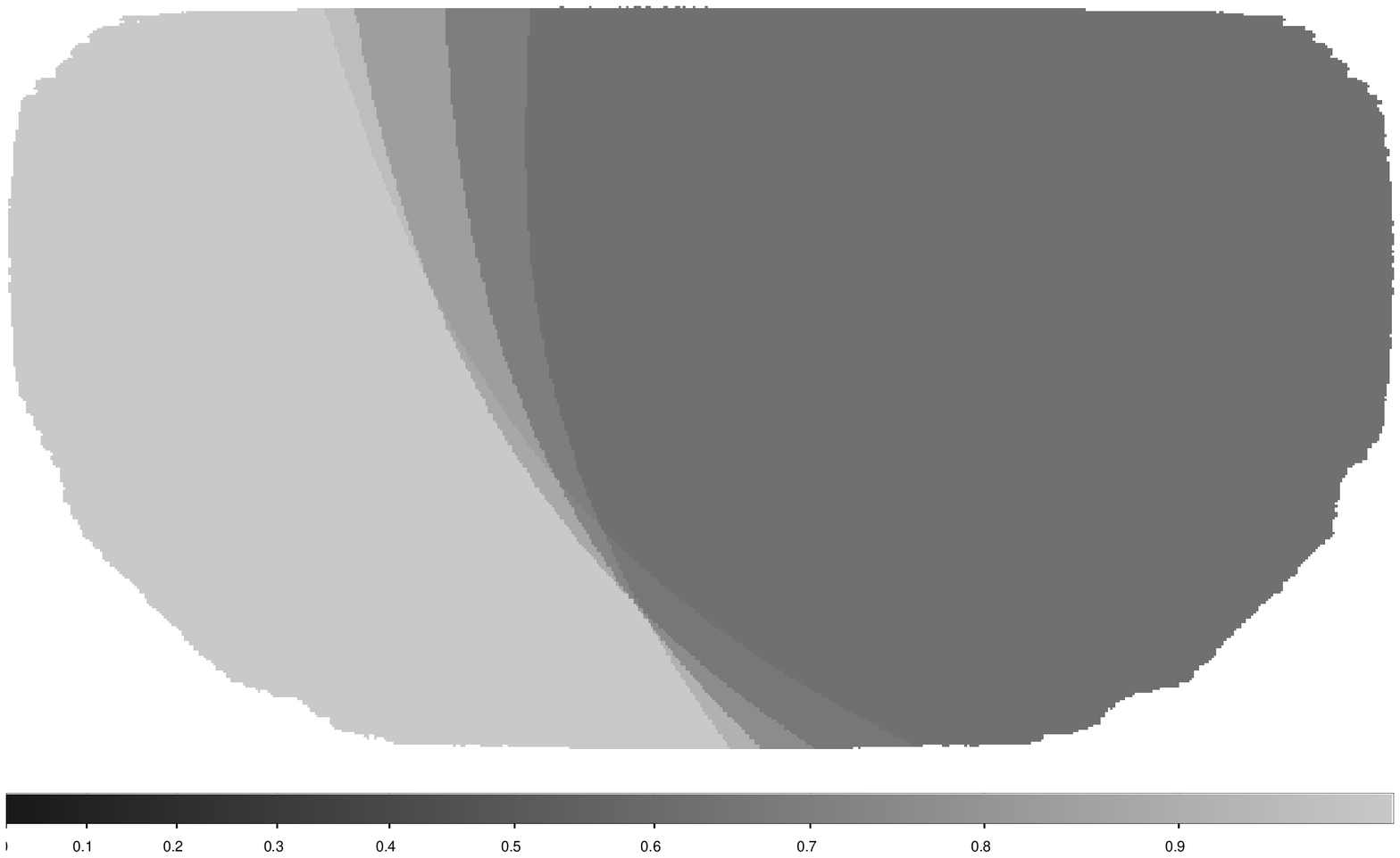}\\
\end{tabular}
  \end{center}
  \caption{
{\bf Left Panel:}
BAT rate, in the 18--20\,keV range, as a function of the elevation angle 
above the  Earth horizon. For ELV $<$30--60$^{\circ}$, BAT starts to experience
Earth occultations. The decrease in rate is expected to vary linearly with
the occulted solid angle if Earth emission is negligible. Note that this
graph is not a light curve as each data point is a separate (in time)
 300 s observation.
{\bf Right Panel:}
Example of deep Earth occultation of the BAT FOV. The black area is
the region of the BAT FOV which is completely occulted by the Earth
during the 300 s observation; the lighter gray is the un-occulted part of the 
FOV, while the region in between the black and the lighter gray is partially
occulted due to the spacecraft movement in the 300\,s. The colorbar shows the
fractional time a given sky pixel is unocculted.
}
  \label{fig:occ}
\end{figure*}

%%%
%%%
As the Earth partially occults the FOV, BAT registers a sharp decrease in rate
due to occultation of the CXB emission. This is especially evident below 40\,keV
where the CXB radiation dominates above the Earth's atmospheric components. 
The left panel of Fig.~\ref{fig:occ} clearly shows  the drop 
in rates caused by the  occulting Earth; between 18--20\,keV 
the rates drop by a factor 3.5 when $\sim$ 60\% of the BAT FOV is occulted
(as shown in the right panel of \ref{fig:occ}).
Thus, the Earth occultation  can be used to measure the CXB emission by means
of the depression caused in the BAT rates. 
Unfortunately, the Earth is not only a passive occulter, 
but also an active emitter. The Earth is a powerful 
source of X- and gamma-rays  due to cosmic ray bombardment of its 
atmosphere \citep[see][]{petry05,sazonov07b}. 
This radiation is usually referred to
as albedo, and it is discussed briefly in the next section.

%%%%%%%%%%%%%%%%%%%%%%%%%%%%%% --------- Fig 5

%
%%%%%%%%%%%%%%%%%%%%%%%%%%%%%%%%%%%%%%%%%%%%%%%%%%%%%%%%%%%%%%%%%%%%%%%%%%
%
%             Atmospheric Albedo
%
\subsection{Atmospheric Albedo Gamma-rays}\label{sec:earth}
The atmospheric Albedo flux is produced by cosmic ray interactions
in the Earth's atmosphere. Hadronic interaction with
atmospheric nuclei of the incident
cosmic rays  leads to the production of an electromagnetic and nuclear
cascade with muons, nuclear fragments, and other
hadrons. 
Gamma-rays above 50 MeV are produced mainly by the decay of mesons, while
at X-ray energies the main source can be attributed to bremsstrahlung 
from secondary electrons.

Measurements of the X-ray albedo radiation are reported in \cite{schwartz74}
in the 1--100\,keV energy range and by \cite{imhof76} above 40\,keV. 
The Albedo spectrum  measured by \cite{schwartz74} shows a cut-off
below 30\,keV, probably due to self-absorption of the radiation emitted
from the inner layers of the atmosphere and  a progressive flattening around
40\,keV. Above 40\,keV the Albedo emission decreases as
a power law with photon index of $\sim$1.4-1.7. 
This power-law behavior is confirmed by other experiments 
\cite[e.g.][]{schoenfelder80,gehrels92} and by a recent Monte Carlo 
simulation of the hard X-ray emission of the 
Earth's atmosphere \citep{sazonov07b}.
However the absolute normalization of the
observed Earth spectrum depends on the altitude and the inclination of the 
satellite's orbit and on the solar cycle.

%
%%%%%%%%%%%%%%%%%%%%%%%%%%%%%%%%%%%%%%%%%%%%%%%%%%%%%%%%%%%%%%%%%%%%%%%%%%
%
%                 Method of analysis
%
\subsection{Method of analysis}
The rate $R(E)_i$  measured at energy $E$ 
by BAT in a given observation, during which 
the Earth is in the FOV, can be described as:
\begin{equation}\label{eq:occ1}
R(E)_i = I(E)_i - \bar{\Omega}_i\cdot[ R(E)_{CXB,i} - R(E)_{Earth,i}]
\end{equation}

where the subscript $i$ refers to the $i-th$ observation, $\bar{\Omega_i}$ 
is the ``effective''
solid angle occulted by the Earth, $I(E)$ is the total background,
and $R(E)_{CXB}$ and $R(E)_{Earth}$
are the CXB and the Earth emission respectively.

The observations we are dealing with are generally non-contiguous,
and thus  all changes in the instrument configuration 
(e.g. number of working detectors) must be taken into account. 
We do this by computing  the ``effective'' solid angle 
occulted by the Earth for each observation.
This is defined as:
\begin{equation}\label{eq:effomega}
\bar{\Omega}_i = \sum_{j=0}^{N_p} \omega_j \cdot (1-\Delta T^{Frac}_j) \cdot V^i_j
\end{equation}
where $N_p$ is the total number of sky pixels,
 $\Delta T^{Frac}_j$ is the fractional exposure time\footnotemark{}
\footnotetext{The fractional exposure time is the fraction 
of the exposure time the sky pixel is unocculted. Thus, it
varies from 0 to 1 for completely occulted and unocculted pixels respectively.}
 a sky pixel
of solid angle $\omega_j$ is unocculted and $V^i_j$ is the 
vignetting affecting that sky pixel during the $i$-th observation.

Equation \ref{eq:occ1} shows the ``degeneracy'' problem which limits
the Earth occultation technique when used to determine the CXB emission.
Indeed, the measured depression of the rates with respect to the normal
sky intensity level are a measurement of the difference 
of the CXB intensity and the Earth's atmospheric emission. 
Following the notation of Equation \ref{eq:occ1}, this can be 
expressed as $R(E)_i=R(E)_{CXB,i}-R(E)_{Earth,i}$.
We adopt here an approach similar to the one 
of \cite{churazov07} and \cite{frontera07},
which consists of deriving the ``difference'' ({\it ON - OFF}) 
spectrum  and fitting it with a-priori spectral models.

%%%%%%%%%%%%%%%%%%%%%%%%%%%%%% --------- Fig 6
\begin{figure}[ht!]
  \begin{center}
  	 \includegraphics[scale=0.80]{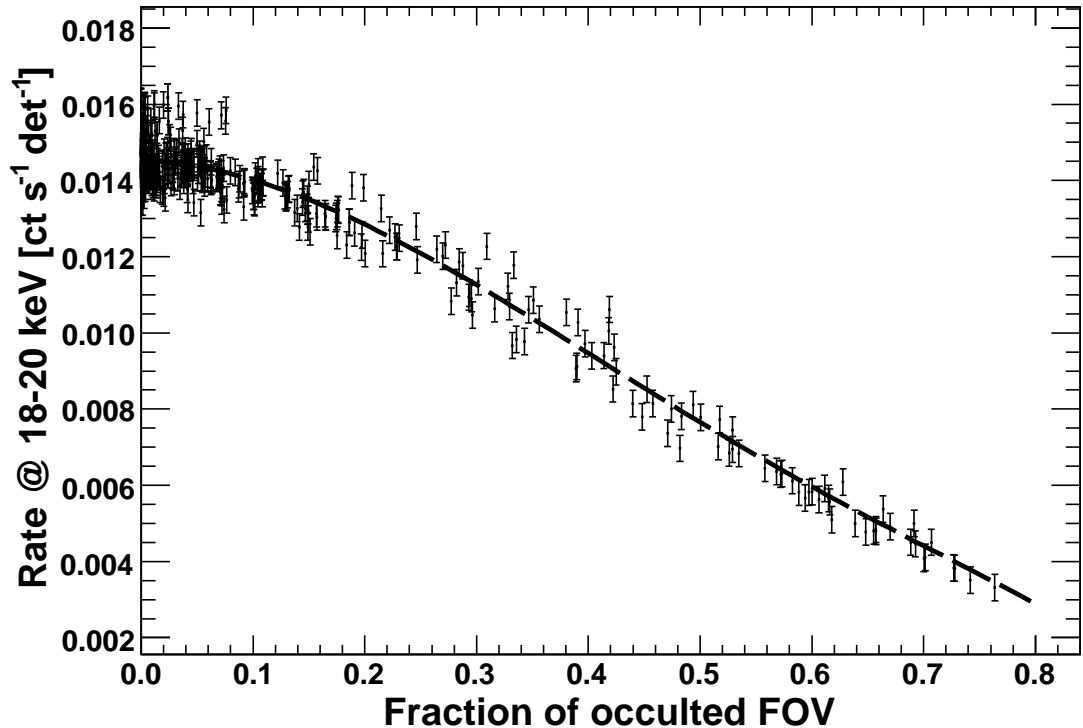}
  \end{center}
  \caption{Fit example for the 18-20 keV energy channel. 
The long dashed line is the best fit to the data. Note that 
the occultation of the CXB emission produces  a very strong signal
reducing the background rate by $\sim$75\%.
 \label{fig:fit_ex}}
\end{figure}

\begin{figure}[ht!]
  \begin{center}
  	 \includegraphics[scale=0.80]{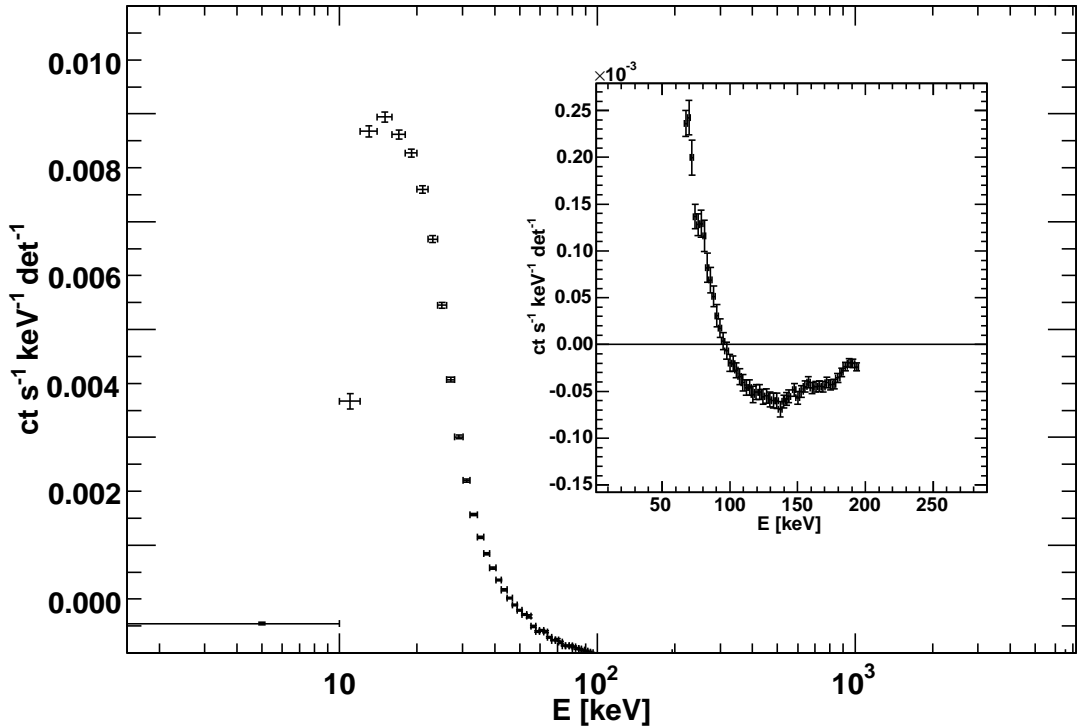}
  \end{center}
  \caption{The {\it ON-OFF} difference spectrum obtained fitting Equation~\ref{eq:occ1}
to each energy channel as shown in Fig.~\ref{fig:fit_ex}. The inset shows
the negative part of the spectrum. Above 100\,keV the albedo spectrum 
dominates the CXB emission.
 \label{fig:diff_raw}}
\end{figure}

The difference spectrum is derived fitting Equation~\ref{eq:occ1}
to each energy channel (an example is shown in
Fig.~\ref{fig:fit_ex}). In all these independent fits
the two parameters ($I(E)_i$ and $R(E)$) are left unconstrained.
Moreover, in order to avoid contamination 
by the Galactic Ridge emission we used
only occultation episodes at Galactic latitude larger than 20$^{\circ}$.
The difference spectrum is shown in Fig.~\ref{fig:diff_raw}.
However, before describing the spectral fit we discuss in detail
the sources of systematic uncertainties affecting our analysis.

%%%%%%%%%%%%%%%%%%%%%%%%%%%%%% --------- Fig 7

%%%%%%%%%%%%%%%%%%%%%%%%%%%%%%%%%%%%%%%%%%%%%%%%%%%%%%%%%%%%%%%%%%%%%%%%%%
%%%%%%%%%%%%%%%%%%%%%%%%%%%%%%%%%%%%%%%%%%%%%%%%%%%%%%%%%%%%%%%%%%%%%%%%%%
\subsubsection{A Note on the ``Degeneracy''  Problem}\label{sec:deg}

The ``degeneracy'' problem (i.e. the fact that the CXB and the 
albedo emissions leave a similar signature during the 
occultation of the BAT FOV) might in principle be
alleviated modeling the albedo emission of the Earth.
This involves modeling the emission as a function of the cut-off
rigidity of the visible disk as well as the (reasonably) expected non-uniformity
of the albedo emission (i.e. limb or disk brightening effects).

Indeed, since the albedo emission is generated at different cut-off rigidities,
with respect the local rigidity of the satellite, one might reasonably
expect that patches of the disk located at lower cut-off rigidities
emit a larger X-ray flux. This information might be used to disentangle
the albedo from the CXB signal.
Moreover, the Earth is known to be a non-uniform emitter at MeV and GeV 
energies. Both COMPTEL and EGRET \citep{schoenfelder80,petry05} have shown
that the Earth exhibits a bright limb. Thus, also this information might be used
to model the expedected emission.

However, we note that a few factors limit the application, in this analysis,
 of the modeling described above. The limitations come from the fact that
this analysis is entirely based on survey data. As explained in 
$\S$~\ref{sec:cuts}, the typical integration time for the survey is 300\,s.
Thus, we do not have a time-resolved monitoring (e.g. 1\,s time resolution) of the
transit of the Earth across the BAT FOV, but only 300\,s snapshot observations with
different level of occultations. Moreover, since the Earth is moving in
the FOV within these 300\,s, all the physical quantities (e.g. cut-off rigidity,
fraction of the occulted FOV, etc.) are necessarily averaged over this time.
Another limitation is due to the fact that the observations used here are not
contiguous in time, but separated by weeks or months. Thus, the changing background
conditions limit the precision of this analysis (as also shown in 
$\S$~\ref{sec:ratevar}). 

These facts limit the usage of a precise modeling of the Earth albedo emission
which would allow to disentangle the albedo and the CXB signal without assuming 
a-priori spectral templates. A dedicated Earth observation
with BAT in 'burst' mode (i.e. event-by-event mode) would not only allow
to overcome the problems shown above, but also would extend the energy
range of the measurement up to 350\,keV (instead of 200\,keV) and would also reduce
the systematic uncertainties of the measurement to those related to the
instrumental response only (see $\S$~\ref{sec:err}).

%%%%%%%%%%%%%%%%%%%%%%%%%%%%%%%%%%%%%%%%%%%%%%%%%%%%%%%%%%%%%%%%%%%%%%%%%%
%%%%%%%%%%%%%%%%%%%%%%%%%%%%%%%%%%%%%%%%%%%%%%%%%%%%%%%%%%%%%%%%%%%%%%%%%%
\section{Analysis of the  Uncertainties}\label{sec:err}

%
%%%%%%%%%%%%%%%%%%%%%%%%%%%%%%%%%%%%%%%%%%%%%%%%%%%%%%%%%%%%%%%%%%%%%%%%%%
%
%                  ERROR Analysis
\subsection{Rate Variation}\label{sec:ratevar}

The rate-rigidity graphs (examples are shown in Fig.~\ref{fig:rate_rig})
show a scatter in the rate around the best fit which is generally larger
than the statistical errors. This scatter is 
due to unknown effects.
The pointing directions, the solar cycle, the
spacecraft orientation with respect  to the 
 Earth and the Sun could be at the origin of
this scatter which has an amplitude of less than 10\%.
We modeled the scatters
as a Gaussian distribution such that the  1\,$\sigma$ width 
of this distribution gives
for each energy channel an estimate of the total (statistical plus
systematic) error of the extrapolated rates. This constitutes
the baseline uncertainty of this analysis, and it is propagated  throughout
all the further steps.

%%%%%%%%%%%%%%%%%%%%%%%%%%%%%%%%%%%%%%%%%%%%%%%%%%%%%%%%%%%%%%%%%%%%%%%%%%%%
\subsection{Uncertainties connected to imprecise attitude determination}
\begin{figure*}[ht!]
  \begin{center}
  \begin{tabular}{cc}
    \includegraphics[scale=0.43]{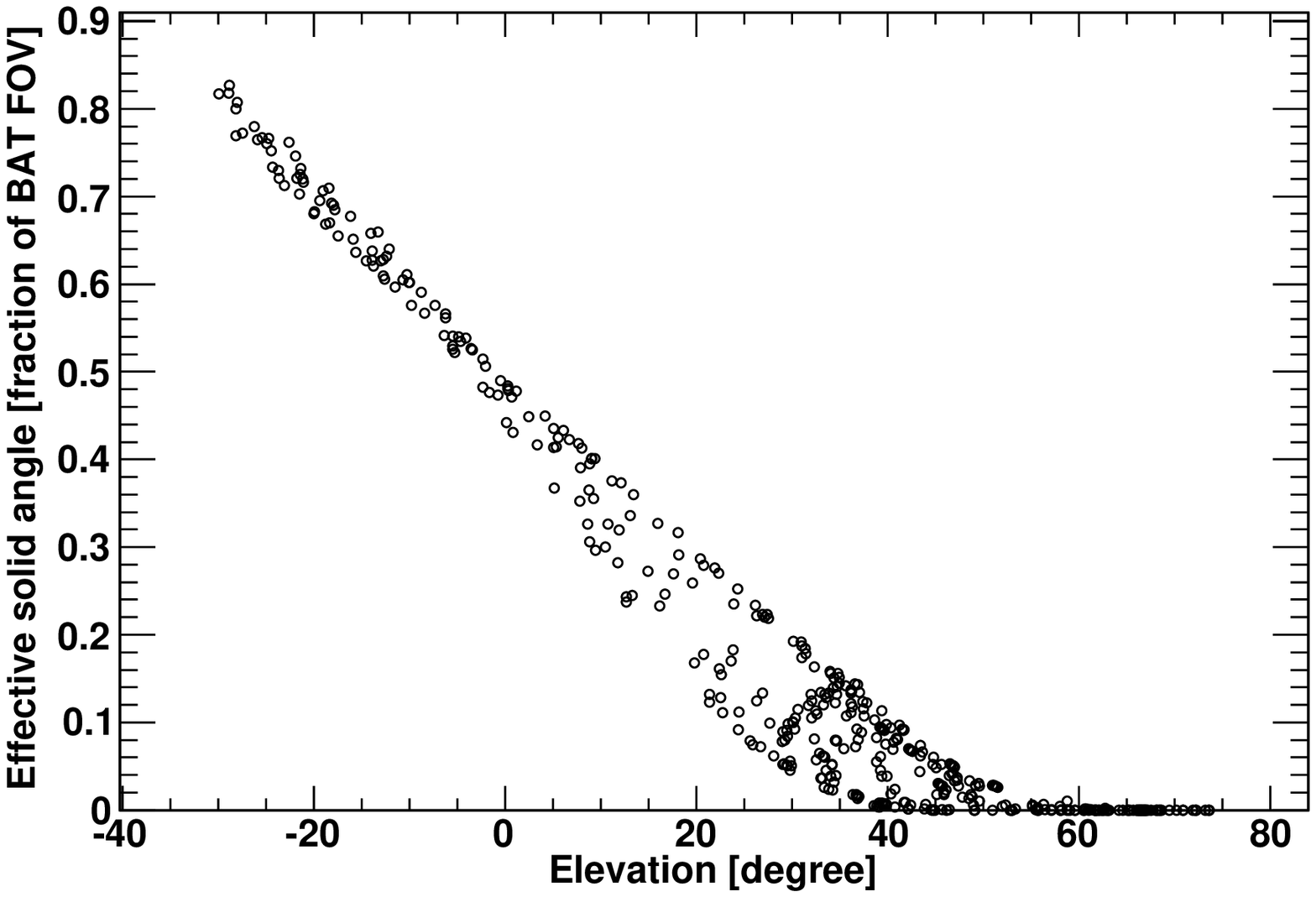}
  	 \includegraphics[scale=0.43]{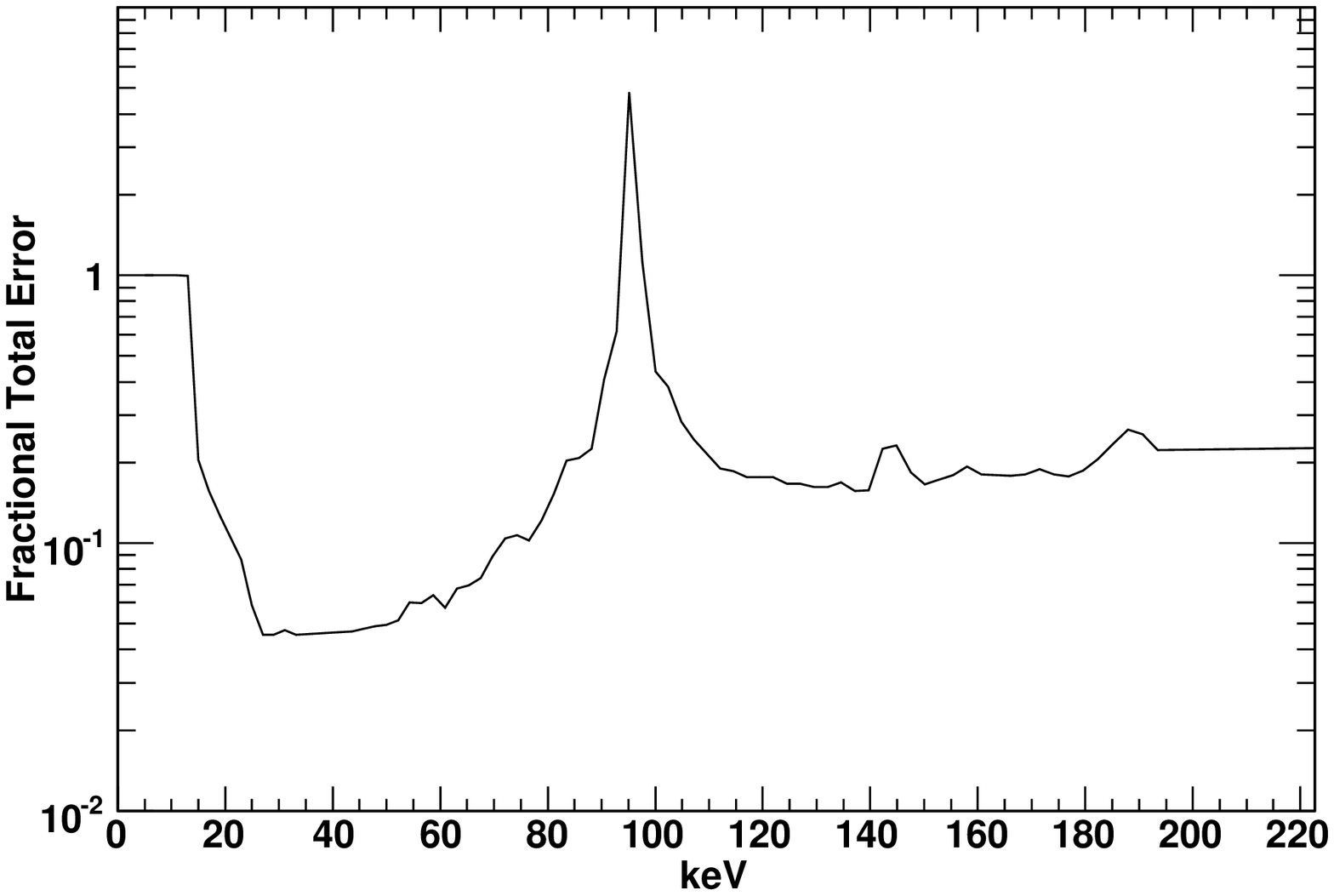}
\end{tabular}
  \end{center}
  \caption{
{\bf Left Panel:}
Variation of the fractional effective solid angle as a function
of elevation (distance of the pointing direction with respect
the Earth limb).
{\bf Right Panel:}
Fractional 1\,$\sigma$ total uncertainty as a function of energy. The fractional 
uncertainty includes all the error estimates outlined
in $\S$~\ref{sec:err}.
The total error reaches its minimum  of 4--5\,\%
at the peak of the CXB emission. The peak around 100\,keV is artificial
and  corresponds to the change of sign of the difference spectrum.
Uncertainties above 100\,keV are $\sim$20\,\% and primarily systematic 
in origin.
 \label{fig:errors}}
\end{figure*}

During safe-mode operations, attitude determination relies
on the magnetometers and Sun sensor. The derived  attitude
solution has a precision of the order of $\sim$degree.
This is confirmed by the 
analysis of sources detected during safe-mode pointings,  which shows that 
the attitude differs from the nominal pointing direction by 1--2 degrees.
The effective solid angle, computed in Eq.~\ref{eq:effomega}, is a 
slowly-varying function per degree of occultation.
As shown in the right panel of Fig.~\ref{fig:errors}, 
the fractional effective solid angle
can be approximated by a straight line
with a slope of 0.010\,deg$^{-1}$ in the 0.4--0.8 range of 
fractional occulted FOV.
This means
that an error of (at most) 2 degrees in the attitude determination
translates into an uncertainty of $\sim$2\% in the determination of 
the occulted portion of BAT FOV. This additional systematic
uncertainty is taken into account in our analysis.

%%%%%%%%%%%%%%%%%%%%%%%%%%%%%% --------- Fig 8

%%%%%%%%%%%%%%%%%%%%%%%%%%%%%%%%%%%%%%%%%%%%%%%%%%%%%%%%%%%%%%%%%%%%%%%%%%%%
\subsection{Uncertainties of  the BAT instrumental response}
\label{subsec:sysbat}
The BAT is a well calibrated instrument.  However, given the very large
FOV and the uncertainty in the modeling of spacecraft materials, the
Crab Nebula is detected with slightly different spectral parameters across
the FOV. To cope with this uncertainty, users are encouraged, when performing 
spectral fitting, to use 
a vector of energy-dependent 
systematic errors\footnotemark{} which allows a unique spectral fit
to the Crab Nebula wherever in the FOV.
\footnotetext{A detailed discussion is reported in http://swift.gsfc.nasa.gov/docs/swift/analysis/bat\_digest.html.}
In this analysis, we account for such systematic errors, which fortunately have their minimum ($\sim$4\%) in the 20-80 keV band.

%%%%%%%%%%%%%%%%%%%%%%%%%%%%%%%%%%%%%%%%%%%%%%%%%%%%%%%%%%%%%%%%%%%%%%%%
%%%%%%%%%%%%%%%%%%%%%%%%%%%%%%%%%%%%%%%%%%%%%%%%%%%%%%%%%%%%%%%%%%%%%%%%
\section{Instrumental response to a diffuse source}\label{sec:resp}
The BAT response was developed by characterizing individual CdZnTe 
detector pixels, and by modeling the absorption and modulation of 
the coded aperture mask, then finally verifying by Monte Carlo 
simulations with radio active sources \citep{sato05}.
However, since there remained uncertainty in response to continuum emissions,
the response was adjusted to fit the Crab nebula spectrum$^4$. 
The BAT Crab spectrum can be described as: 
$dN/dE = 10.40\ E^{-2.14}$ photons cm$^{-2}$ s$^{-1}$ keV$^{-1}$.
The values of normalization and photon index are well within
those used by most of the X-ray missions in a similar band
\citep[see ][ for a review of Crab Nebula spectral parameters]{kirsch05}.

However, the analysis of a diffuse source (as the CXB) presents some
differences with respect to the study of point-like objects.
Indeed, the spectrum of a point-like source is modulated by the coded mask pattern.
Thus, the indirect unmodulated component which is scattered by the materials 
of the BAT instrument and of the satellite can  be  eliminated.
Accordingly, the official response generator, {\it batdrmgen}, 
part of the standard BAT software,
produces a response only for the direct component.
However, the CXB, the subject of this paper, is seen as a diffuse emission and cannot be modulated by the coded mask pattern.
We therefore utilized the Monte Carlo simulator to generate a 
more accurate response for a diffuse emission 
taking into account the scattered component\footnotemark{} 
as well as the effect of isotropic illumination of the BAT instrument.
This simulator is the one used to verify the response on the ground,
but the same corrections to fit the Crab spectrum are also applied.
\footnotetext{The so called 'un-coded' (or scattered) component comprises
all those events which scatter in the satellite structure and produce
a detectable signal in the BAT array.}

%
%%%%%%%%%%%%%%%%%%%%%%%%%%%%%%%%%%%%%%%%%%%%%%%%%%%%%%%%%%%%%%%%%%%%%%%%%%%%%
%%%%%%%%%%%%%%%%%%%%%%%%%%%%%%%%%%%%%%%%%%%%%%%%%%%%%%%%%%%%%%%%%%%%%%%%%%%%%
%    Results
%
\section{Results of the analysis}\label{sec:results}
%%%
In this section we present the main results of the analysis:
the CXB and the Earth's atmosphere spectra. 
All the uncertainties described in  $\S$~\ref{sec:err}
were added in quadrature to form the total uncertainty. 
The  dependence of the total uncertainty
with energy is shown in 
Figure~\ref{fig:errors} (right panel). The total uncertainty reaches its
minimum value of 4--5\,\% at the peak of the CXB spectrum.

In this section, all quoted errors on spectral parameters
 are 90\% confidence for one interesting parameter.

%%%%%%%%%%%%%%%%%%%%%%%%%%%%%%%%%%%%%%%%%%%%%%%%%%%%%%%%%%%%%%%%%%%%%%%%%%%%%
\subsection{Spectral fitting}
The {\it ON-OFF} difference spectrum is folded in XSPEC \citep{arnaud96} 
with the proper instrumental
response for a diffuse source.  The model we used for the fit is the 
difference between the CXB and the albedo spectra.
For the CXB spectrum we employ Equation~\ref{eq:gruber}.
For the albedo spectrum we use a jointly smoothed double power-law
of the form:
\begin{equation}\label{eq:joint}
\frac{dN}{dE} = \frac{C}{(E/E_{b})^{\Gamma_1} +(E/E_{b})^{\Gamma_2} }\ \ \ \ 
[\rm{photons\, cm^{-2}\,s^{-1}\,sr^{-1}}]
\end{equation}
%%%%%%%
%%%%%%%

where $\Gamma_1$ and $\Gamma_2$ are the two spectral indices and $E_{\rm b}$
is the break energy.
This functional form reproduces well the atmospheric component
with its declinde at low energy, bump around 30--40\,keV 
and a hard spectral
index at higher energies \citep{schwartz74,imhof76,gehrels85,frontera07}.
Recent Monte Carlo simulations of the Earth emission \citep{sazonov07b}
show that Equation~\ref{eq:joint} is a very good approximation of
the Earth emission below 300\,keV.
We fix the values of the spectral indices and break energy
at those suggested by \cite{sazonov07b}, i.e. 
$\Gamma_1$=-5 and $\Gamma_2$=1.4 and $E_{\rm b}$=44\,keV.
Thus,  free parameters of our first fit are only
the normalizations of Equations~\ref{eq:gruber} and \ref{eq:joint},
respectively. The fit is poor, however, with a $\chi^2$ of $\sim220$ for
75 degrees of freedom.
Adding free parameters for the high energy spectral index $\Gamma_2$
and the break energy $E_{\rm b}$ improves the fit 
($\chi^2_{red}$=121.8/73). The F-test confirms that the improvement
is very significant (F-test probability of 4.8$\times 10^{-10}$).
Adding another free parameter for 
the low energy spectral index $\Gamma_1$ of the
albedo does not improve the fit. Indeed, below 40\,keV the spectrum
is completely dominated by the CXB emission, and thus it is not
possible to constrain this parameter.  
Choosing as free variables the parameters of the 
Earth emission instead of those of the CXB spectrum is well motivated:
there are indications \citep{schwartz69,gehrels85,frontera07}
that the high-energy spectral index of the albedo emission
might be steeper than the classical value of 1.4.
On the other side, the  formula shown in Eq.~\ref{eq:gruber} \citep{gruber99}
is a good representation of the broad band CXB spectrum.

\begin{figure}[ht!]
  \begin{center}
  	 \includegraphics[scale=0.6,angle=270.]{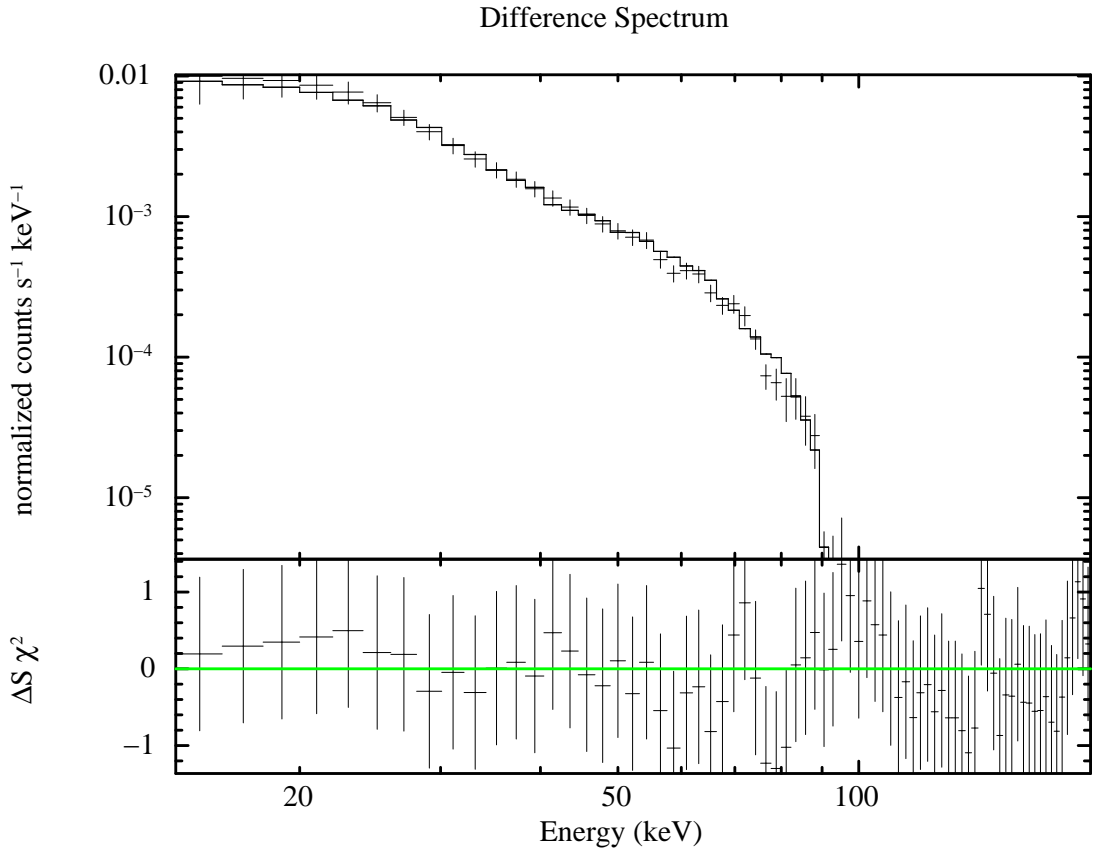}
  \end{center}
  \caption{Best fit to the {\it ON-OFF} 
difference spectrum. The model used is the 
difference between Equations ~\ref{eq:gruber} and \ref{eq:joint}.
Around 100\,keV, the data become negative (as shown in Fig.~\ref{fig:diff_raw}).
 \label{fig:residuals}}
\end{figure}

Our best-fit parameters (with 90\,\% CL errors) for the albedo spectrum are:
$\Gamma_2$=1.72$\pm0.08$, $E_{\rm b}$=33.7$\pm3.5$\,keV
  and $C$=1.48$^{+0.6}_{-0.3} \times 10^{-2}$.
The normalization, and its 90\,\% CL error,
 of the CXB as measured by BAT 
with respect Equation~\ref{eq:gruber}  is 1.06$\pm0.08$. 
This error includes also the change
by $\pm1$ in the low-energy spectral index ($\Gamma_1$) 
of the albedo emission. The CXB intensity in the 20--50\,keV band
is 6.43($\pm0.20$)$\times10^{-8}$ \,erg cm$^{-2}$ s$^{-1}$ sr$^{-1}$.
Figure~\ref{fig:residuals} shows the  best fit 
and its residuals  to the difference spectrum.

%%%%%%%%%%%%%%%%%%%%%%%%%%%%%% --------- Fig 9

%%%%%%%%%%%%%%%%%%%%%%%%%%%%%%%%%%%%%%%%%%%%%%%%%%%%%%%%%%%%%%%%%%%%%%%%%%%%%
%%%%%%%%%%%%%%%%%%%%%%%%%%%%%%%%%%%%%%%%%%%%%%%%%%%%%%%%%%%%%%%%%%%%%%%%%%%%%
\section{Alternative measurement of the X-ray background spectrum}\label{sec:quad}

\begin{figure*}[ht!]
  \begin{center}
  \begin{tabular}{cc}
    \includegraphics[scale=0.41,clip=true,trim=0 33 0 34]{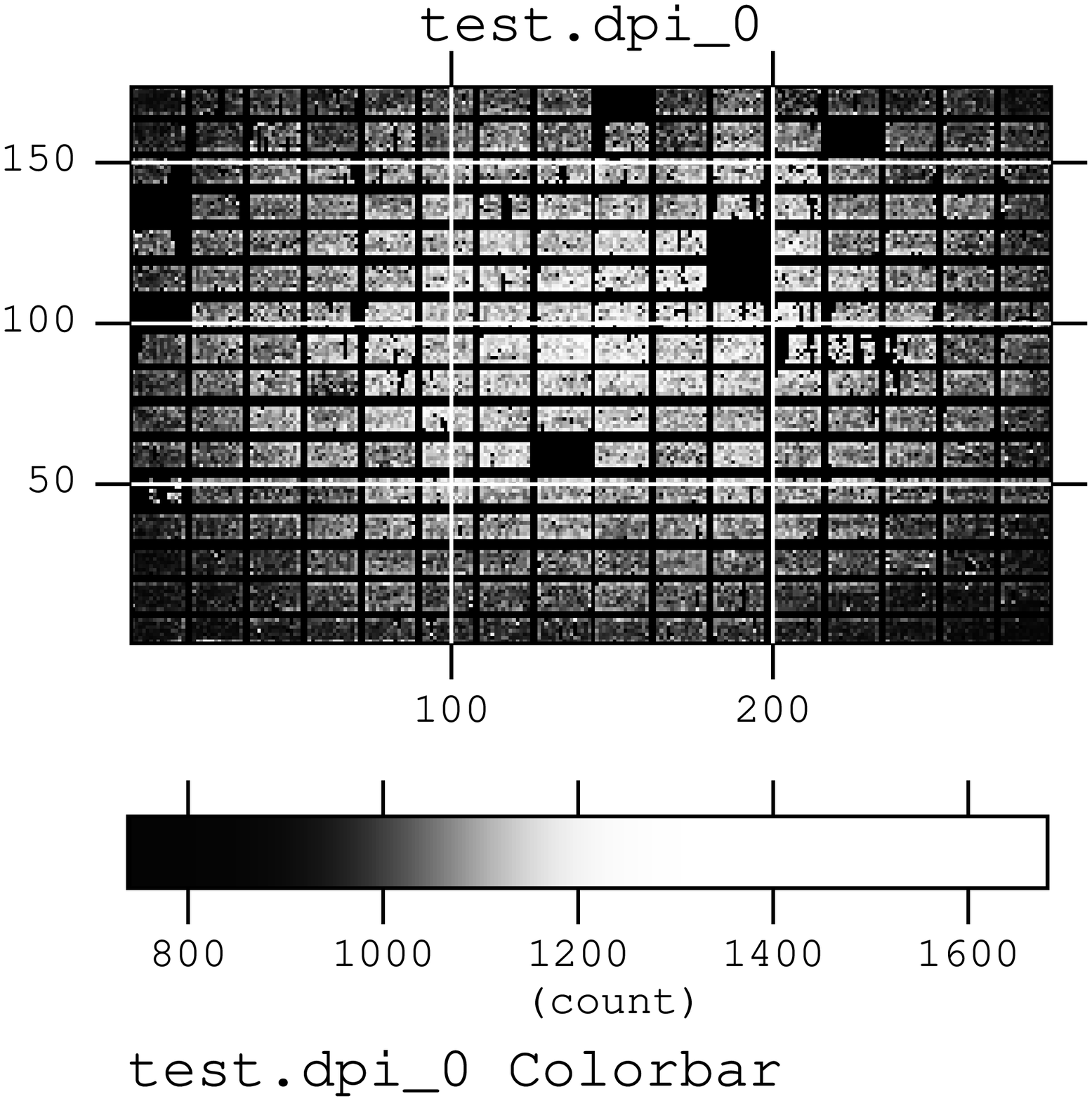}
  	 \includegraphics[scale=0.485,clip=true,trim=0 30 0 34]{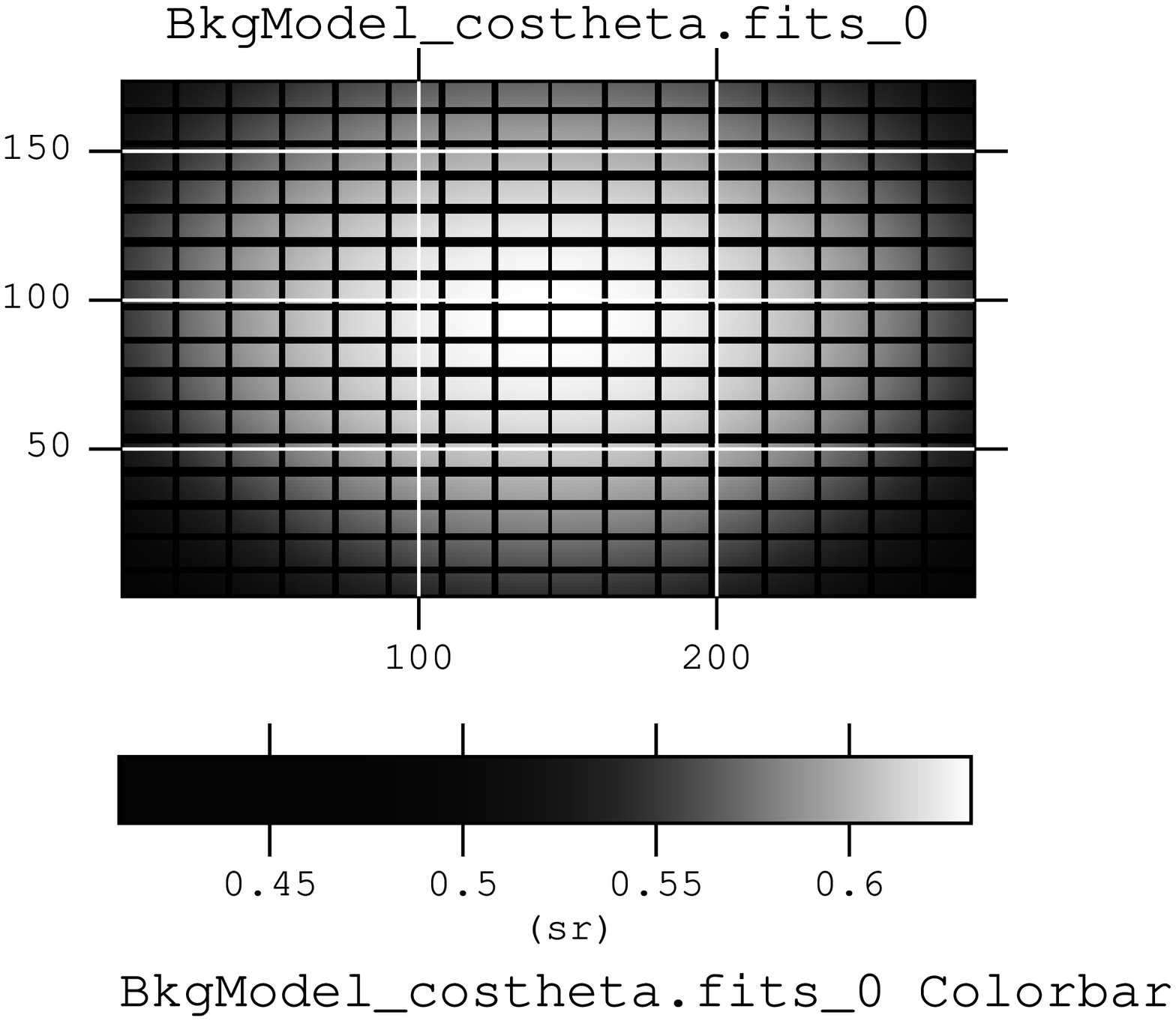}\\
\end{tabular} 
  \end{center}
  \caption{
{\bf Left Panel:}BAT detector plane image in the 15--55\,keV energy channel.
{\bf Right Panel:}Model of the detectors solid angles distribution. 
The similarity of the model with the real data (left panel) is apparent.
 \label{fig:quadbkg}}
\end{figure*}

The BAT in-flight background has a peculiar spatial distribution which
shows larger count rates towards the center of the detector array and smaller
rates towards its edges. This is clearly shown in Fig.~\ref{fig:quadbkg} 
(left panel). 
 In the process of forming sky images, 
the BAT software\footnotemark{} removes this background
component by means of an empirical  bi-dimensional second-order 
polynomial function. \footnotetext{For reference see the description
of the {\it batclean} tool available at 
http://heasarc.nasa.gov/lheasoft/ftools/headas/batclean.html.}

%%%%%%%%%%%%%%%%%%%%%%%%%%%%%% --------- Fig 10

This feature has 
an important  physical meaning. Indeed, it
is the result of a diffuse isotropic source (namely the CXB),
shining through the mask. The peculiar shape of this ``background'' component
arises from the fact that detectors at the edge of the array have a smaller
solid angle of the sky as seen 
through the mask (they see the mask under large angles)
than those at the very center. Given the extent of the BAT mask and array,
this effect is noticeable and significant.

We thus built a simple model, assigning to each detector its geometrical
solid angle through the transparent mask elements. This model is shown 
in Fig.~\ref{fig:quadbkg}.  
We can detect the CXB emission by fitting this model to 
the spatial distribution of the counts in each energy channel. 
However, this approach is valid only  as long as the graded-Z shield remains
opaque to X-ray photons($\sim$50\,keV). 
Indeed, as soon as the shield becomes partially
transparent, the effective detector solid angle increases because of the shield
transparency. Thus, our model becomes inadequate above this energy.

%%%%%%%%%%%%%%%%%%%%%%%%%%%%%%%%%%%%%%%%%%%%%%%%%%%%%%%%%%%%%%%%%%%%%%%%%%%%
%%%%%%%%%%%%%%%%%%%%%%%%%%%%%%%%%%%%%%%%%%%%%%%%%%%%%%%%%%%%%%%%%%%%%%%%%%%%
\subsection{Model fitting}\label{sec:modelfitting}
Among all BAT observations which satisfied the selection criteria
outlined in $\S$~\ref{sec:cuts}, we selected only those ones
which were unocculted by the Earth. 
We then summed all the detector plane histograms\footnotemark{}
\footnotetext{BAT survey data are in the form of 80 channels detector plane 
histograms with a typical exposure time of  300\,s.}
(DPHs)
into a single DPH with an overall exposure of $\sim$1.8\,Ms. Summing
the DPHs of observations with different pointing directions achieves the goal
of smearing the contribution of sources which are below the detection
threshold (8\,$\sigma$ in this case). To each energy channel we fitted a model
which is composed of:
\begin{itemize}
\item a constant term for the edges of the detector modules which register a
higher count rate because of the larger exposed area,
\item the solid angle distribution  model which takes into account the diffuse
flux as seen through the mask,
\item a constant term which takes into account all other background 
components including 
the CR component which penetrates through the shielding.
\end{itemize}

For each energy channel, the fit independently determines the intensity
of the diffuse model. 
Since all the energy-dependent effects (e.g. absorption through
the mask structure and transparency of the lead tiles) are correctly 
taken care of in the instrumental response described in $\S$~\ref{sec:resp},
we normalize our model  (dividing by  the maximum detector solid angle 
$\sim$0.6\,sr) and treat the dispersion of the solid angle
distribution ($\sim$0.034\,sr)  as a fluctuation.
In this way, we make our approach insensitive to the exact computation
of the solid-angle for each detector and at the same time, it allows us to use
the same response matrix developed for the occultation measurement.
We remark that this response matrix is based
on extensive Monte Carlo simulations.

%%%%%%%%%%%%%%%%%%%%%%%%%%%%%%%%%%%%%%%%%%%%%%%%%%%%%%%%%%%%%%%%%%%%%%%%%%%%
%%%%%%%%%%%%%%%%%%%%%%%%%%%%%%%%%%%%%%%%%%%%%%%%%%%%%%%%%%%%%%%%%%%%%%%%%%%%
\subsection{Results}
For spectral fitting we convolved the CXB count rate spectrum with the 
BAT response matrix. For each energy channel, we summed in quadrature
statistical uncertainty,
the uncertainty on the mean
solid angle (see  $\S$~\ref{sec:modelfitting}) 
and the uncertainty due to the BAT response  (see $\S$~\ref{sec:resp}).

\begin{figure}[ht!]
  \begin{center}
  	 \includegraphics[scale=0.6,angle=270.]{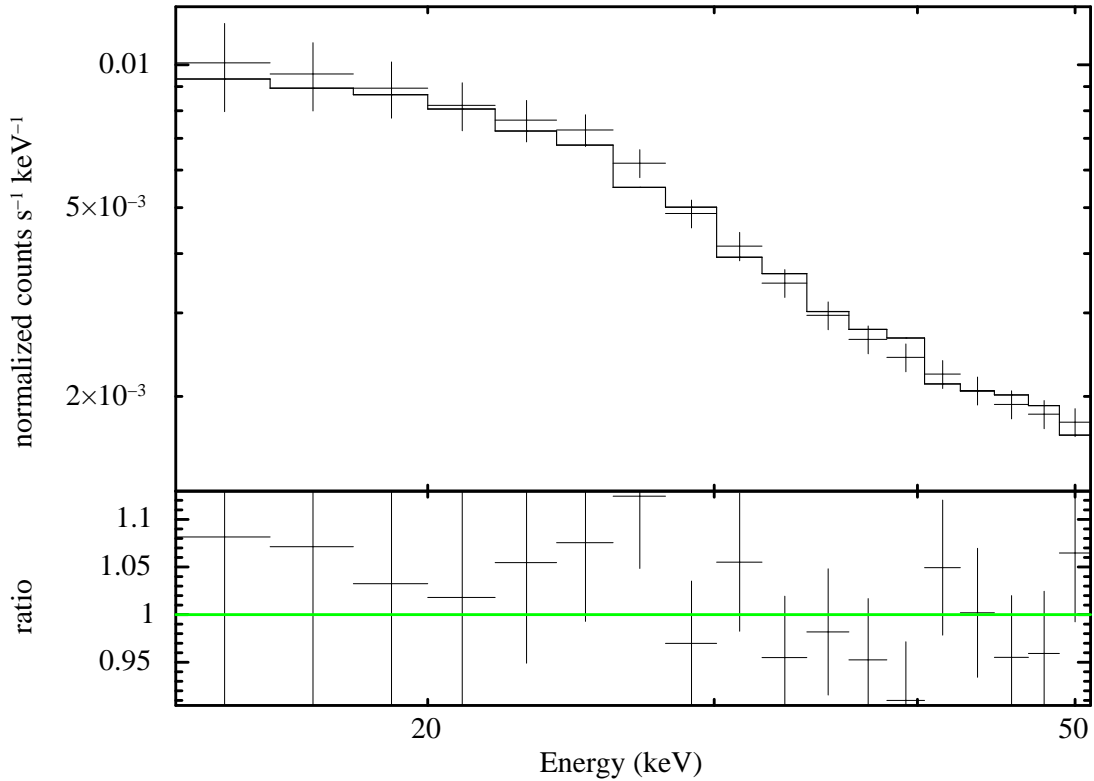}
  \end{center}
  \caption{Best fit to the second CXB measurement using 
Equation~\ref{eq:gruber} with only the overall normalization left
as free parameter. 
 \label{fig:quadfit}}
\end{figure}

\begin{figure}[ht!]
  \begin{center}
  	 \includegraphics[scale=0.89,angle=0.]{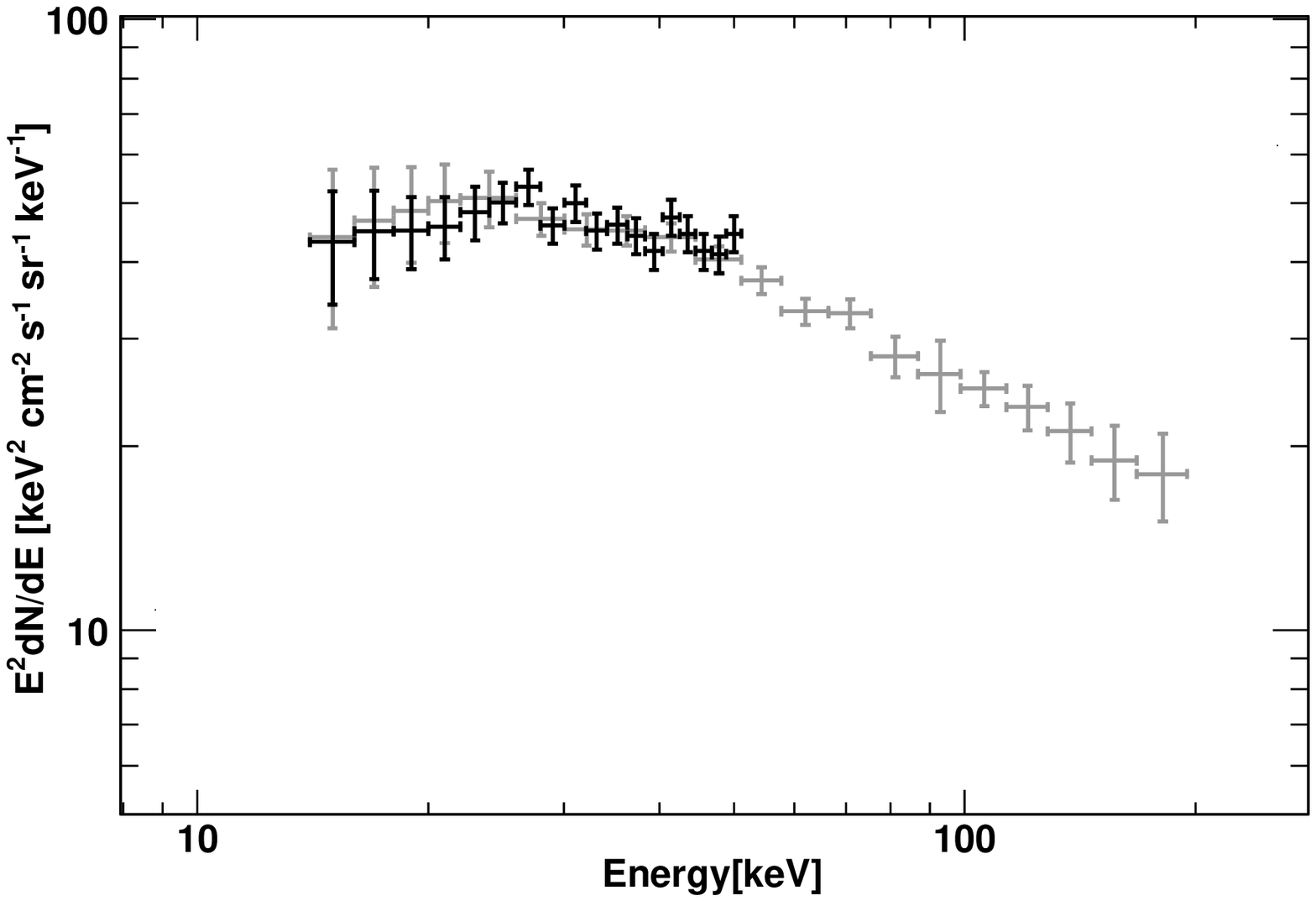}
  \end{center}
  \caption{The two independent measurements 
of the CXB spectrum performed by BAT.
The occultation measurement (gray datapoints) and the measurement
derived using the solid angle distribution (black datapoints) are
in very good agreement.
 \label{fig:cxb_double}}
\end{figure}

A fit  to the data (shown in Fig.~\ref{fig:quadfit})  
allowing only the overall CXB normalization 
(Equation~\ref{eq:gruber})  to vary yields a $\chi^2$ of 18.8 for
17 degree of freedom. The normalization  with respect to
the level of the CXB as measured by \cite{gruber99} is
$1.09^{+0.03}_{-0.03}$. This measurement is in very good 
agreement with the occultation measurement as Fig.~\ref{fig:cxb_double}
shows.

%%%%%%%%%%%%%%%%%%%%%%%%%%%%%% --------- Fig 11

%%%%%%%%%%%%%%%%%%%%%%%%%%%%%% --------- Fig 12

%%%%%%%%%%%%%%%%%%%%%%%%%%%%%%%%%%%%%%%%%%%%%%%%%%%%%%%%%%%%%%%%%%%%%%%%%%%%%
%%%%%%%%%%%%%%%%%%%%%%%%%%%%%%%%%%%%%%%%%%%%%%%%%%%%%%%%%%%%%%%%%%%%%%%%%%%%%
\section{Comparison with previous measurements}\label{sec:disc}
In this section, we compare the CXB and the albedo spectra
with previously available measurements in the same or overlapping
energy bands. For reference, the values of CXB and albedo emissions
as derived by BAT are reported in Table~\ref{tab:cxb}.

\begin{deluxetable}{lccccc}
\tablewidth{0pt}
\tablecaption{Cosmic X-ray Background  and albedo emission intensities
\label{tab:cxb}}
\tablehead{
\colhead{Energy} & \colhead{$\Delta$E}   & 
\colhead{CXB} & \colhead{CXB 1\,$\sigma$ Error} &    
\colhead{Albedo\tablenotemark{a}} & \colhead{Albedo 1\,$\sigma$ Error} \\
%%%%%%%%%  units
\colhead{\footnotesize{keV}} & \colhead{}& 
\colhead{\footnotesize{keV$^2$ cm$^{-2} s^{-1}$ sr$^{-1}$ keV$^{-1}$}} 
& \colhead{}&
\colhead{\footnotesize{10$^{-4}$\,ph cm$^{-2} s^{-1}$ sr$^{-1}$ keV$^{-1}$}} 
& \colhead{} 
}
\startdata 

15.0 & 2.0 & 43.92 & 12.71 & \nodata & \nodata\\
17.0 & 2.0 & 46.82 & 10.35 &\nodata & \nodata\\
19.0 & 2.0 & 48.58 & 8.65 & \nodata & \nodata\\
21.0 & 2.0 & 50.39 & 7.42 & \nodata & \nodata\\
24.0 & 4.0 & 50.96 & 5.28 & \nodata & \nodata\\
28.1 & 4.1 & 47.12 & 2.90 & \nodata & \nodata\\

32.2 & 4.1 & 45.26 & 2.72 & 63.96 & 19.36\\
36.2 & 4.1 & 45.07 & 2.52 & 73.93 & 13.75\\
41.4 & 6.3 & 44.00 & 2.34 & 76.25 & 9.50\\
47.8 & 6.5 & 40.39 & 2.07 & 74.94 & 6.32\\
54.3 & 6.5 & 37.33 & 1.89 & 64.57 & 4.72\\
62.0 & 8.8 & 33.25 & 1.63 & 55.01 & 3.29\\
70.9 & 9.0 & 32.99 & 1.79 & 38.30 & 2.85\\
81.2 & 11.6 & 28.08 & 2.17 & 34.16 & 3.14\\
92.9 & 11.8 & 26.25 & 3.50 & 26.53 & 4.00\\
106.1 & 14.5 & 24.86 & 1.61 & 20.64 & 1.36\\
120.8 & 15.0 & 23.18 & 1.93 & 16.41 & 1.37\\
137.2 & 17.9 & 21.15 & 2.32 & 13.43 & 1.46\\
156.8 & 21.2 & 18.95 & 2.62 & 10.96 & 1.43\\
181.1 & 27.5 & 18.02 & 2.96 & 8.38 & 1.22\\

\enddata
\tablenotetext{a}{In the 14--30\,keV energy range the 1\,$\sigma$
upper limit to the albedo intensity is 6.3$\times 10^{-3}$ ph cm$^{-2}$ s$^{-1}$ sr$^{-1}$  keV$^{-1}$}
\end{deluxetable}

%%%%%%%%%%%%%%%%%%%%%%%%%%%%%%%%%%%%%%%%%%%%%%%%%%%%%%%%%%%%%%%%%%%%%%%%%%%%%
%%%%%%%%%%%%%%%%%%%%%%%%%%%%%%%%%%%%%%%%%%%%%%%%%%%%%%%%%%%%%%%%%%%%%%%%%%%%%
\subsection{The X-ray Background Spectrum}\label{sec:cxbspec}

\begin{figure}[ht!]
  \begin{center}
  	 \includegraphics[scale=0.89,angle=0.]{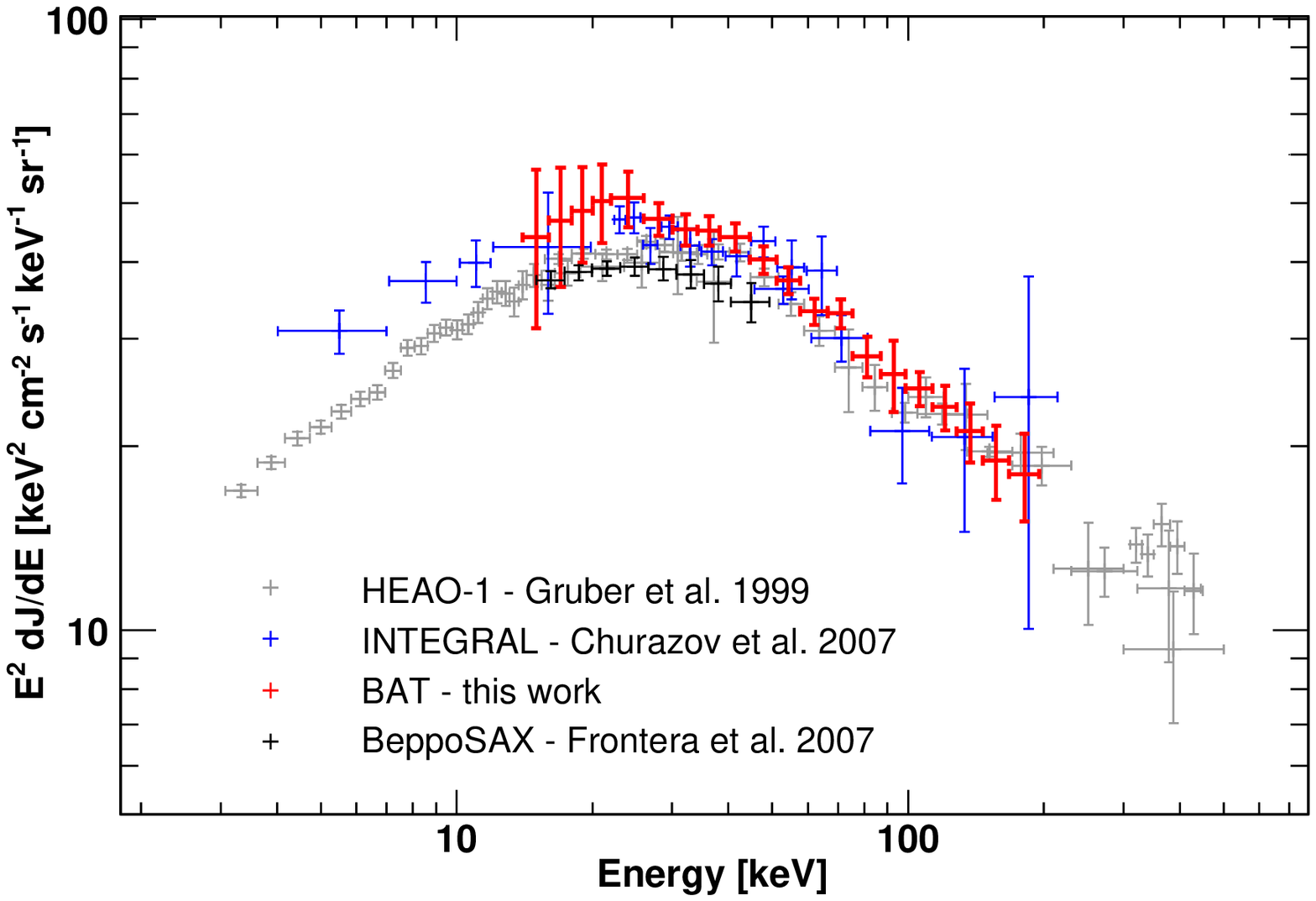}
  \end{center}
  \caption{Comparison of CXB measurements above 20\,keV. 
The BAT spectrum (in red) is in agreement with the HEAO-1 (gray),
{\it INTEGRAL} (blue) and BeppoSAX (black) observations.
For clarity only the BAT occultation measurement is reported.
 \label{fig:comp}}
\end{figure}

Both measurements of the CXB spectrum presented here produce the same
results (within errors) for the normalization of the CXB
intensity at its peak. Combining both measurements
we determine that the CXB intensity at its peak is 8($\pm3$)\,\%
larger than previously measured by HEAO-1 \citep{gruber99}.
We find that the CXB intensity in the 20--50\,keV band
is 6.50$\pm0.15 \times 10^{-8}$\,erg cm$^{-2}$ s$^{-1}$ sr$^{-1}$.
The observed intensity near the peak of the CXB spectrum (expressed
in $\nu$F$_{\nu}$ units)  at 30\,keV
is 46.2 keV$^2$ cm$^{-2}$ s$^{-1}$ keV$^{-1}$ sr$^{-1}$. 
Figure~\ref{fig:comp} shows the comparison of the BAT CXB spectrum
with all other measurements available above 20\,keV. All  measurements
agree well within 10\,\%.
The detailed comparison is reported in  Tab.~\ref{tab:comp}.
It is clear that the scatter in CXB intensities does not depend solely on the 
adopted spectra for the Crab Nebula. Some of the measurements showed
in Tab.~\ref{tab:comp} might still be affected by systematic uncertainty
in the instrumental response used.
To our knowledge, BAT is the only instrument for which a dedicated
instrumental response has been derived and tested for the analysis of the
CXB.

%%%%%%%%%%%%%%%%%%%%%%%%%%%%%% --------- Fig 13

\begin{deluxetable}{lcccc}
\tablewidth{0pt}
\tablecaption{Comparison with previous results\label{tab:comp}}
\tablehead{
\colhead{Instrument} & \colhead{Ref.}   & \colhead{Energy band}    
& \colhead{F$_{{\rm Crab}}$\tablenotemark{a}} & \colhead {I$_{\rm CXB}$\tablenotemark{b}} \\
%%%%%%%%%  units
\colhead{} & \colhead{}& \colhead{keV} & 
\colhead{} & \colhead{}
}
\startdata 
HEAO-1/A2+A4 &  1 & 20--50  & 9.92\tablenotemark{c}  & 6.06$\pm0.06$\\
HEAO-1/A2    &  2 & 20--50  & NA                     & 5.60$\pm0.30$ \\

%RXTE         &  3 & 15--20  & 3.72 & 2.36 $\pm 0.02$   
%&1.93$\pm 0.04$ & 3.05\\ 

BeppoSAX     &  3 & 20--50  & 9.22 & 5.89$\pm 0.19$   \\

%INTEGRAL &  4 & 20--100 & 17.6 & $\sim$ 10.45 \tablenotemark{d}\\
INTEGRAL &  4 & 20--50 & 10.4 & $\sim$6.66 \tablenotemark{d}\\

BAT & this work & 20--50 & 9.42\tablenotemark{e} & 6.50$\pm 0.15$ \\
\enddata
\tablenotetext{a}{Crab flux quoted by the authors expressed 
in 10$^{-9}$\,erg cm$^{-2}$ s$^{-1}$ keV$^{-1}$}
\tablenotetext{b}{Intensity of the CXB quoted by the authors 
in 10$^{-8}$\,keV$^2$ cm$^{-2}$ s$^{-1}$ sr$^{-1}$ keV$^{-1}$}
%%%
\tablenotetext{c}{\cite{gruber99} do not report about their adopted 
Crab spectrum; however the HEAO-A4 spectrum of the Crab Nebula can
be described (below 57\,keV) as 
$dN/dE = 8.76\ E^{-2.0.75}$ photons cm$^{-2}$ s$^{-1}$ keV$^{-1}$
\citep{jung89}.
}
%%%
\tablenotetext{d}{Authors do not give an exact measurement of the CXB flux, but
report that their measurement is $\sim$10\% higher than the \cite{gruber99} 
spectrum.}

\tablenotetext{e}{The value quoted here has to be taken as a reference value.
The systematic uncertainties discussed in $\S$~\ref{subsec:sysbat} allow 
to derive consistent Crab Nebula fluxes across the entire
BAT FOV.}
\tablerefs{(1) \cite{gruber99}; (2) \cite{marshall80}; 
(3) \cite{frontera07}; (4) \cite{churazov07}.
}
\end{deluxetable}

%%%%%%%%%%%%%%%%%%%%%
%%%%%%%%%%%%%%%%%%%%%

\begin{figure}[ht!]
  \begin{center}
  	 \includegraphics[scale=0.89,angle=0.]{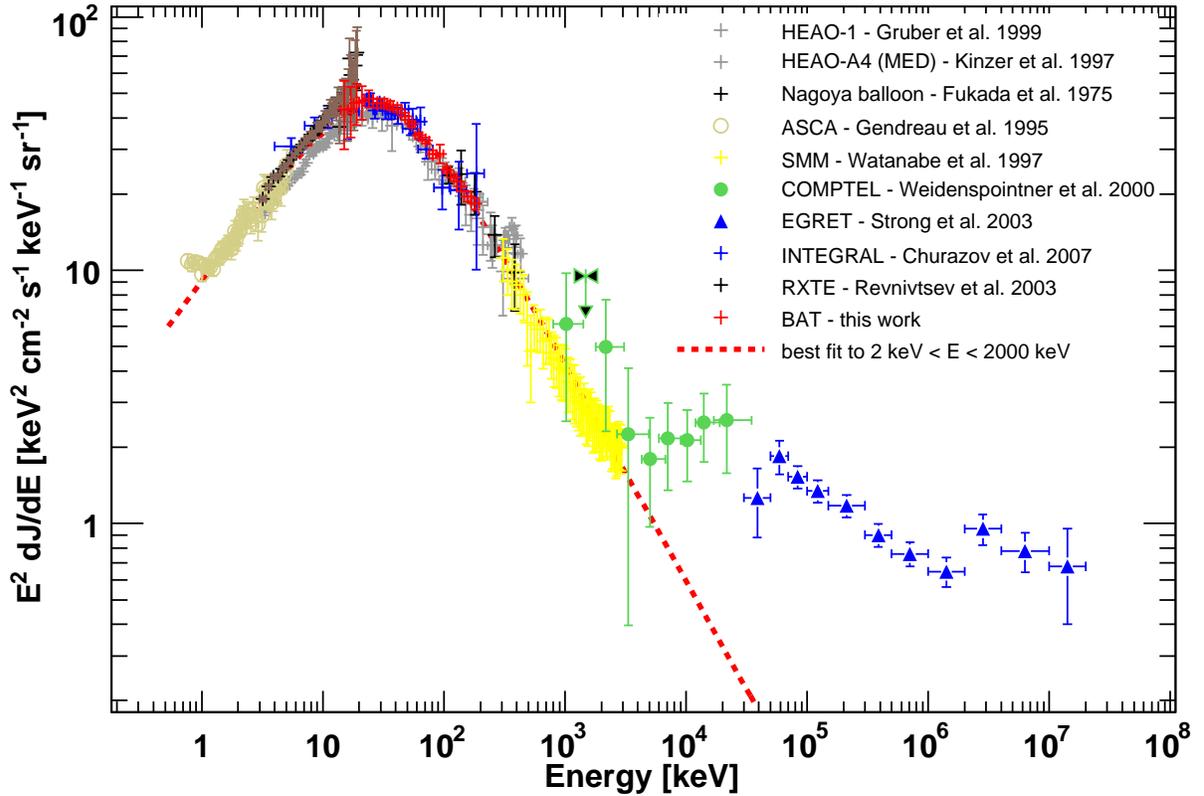}
  \end{center}
  \caption{BAT CXB spectrum compared with previous results. The dashed line
is the best fit to 2\,keV$<$ E $<$ 2000\,keV as reported in the text.
 \label{fig:xall}}
\end{figure}

Fig.~\ref{fig:xall} shows a compilation of the X- and gamma-ray diffuse
backgrounds from keV to GeV energies. In addition to the work of 
\cite{gruber99}, we show  SMM (MeV)  data \citep{watanabe97}, and 
 {\it COMPTEL} and {\it EGRET} data in a recent revision \citep{weidenspointner00,strong04}.
In particular, the new analysis of {\it EGRET }data shows that the 
validity range of the \cite{gruber99} formula (Eq. \ref{eq:gruber})
is now restricted to  3\,keV$<$ E $<$ 1\,MeV.

We find that a good description of the available data in the 
2\,keV -- 2\,MeV range is achieved using a smoothly-joined
double power-law of the form:
\begin{equation}
E^2 \cdot \frac{dN}{dE} = E^2\cdot \frac{C}{(E/E_{\rm B})^{\Gamma_1} +(E/E_{\rm B})^{\Gamma_2}}\ \ \ \ [{\rm keV}^2 {\rm photons\ cm}^{-2} {\rm s}^{-1} {\rm sr}^{-1} 
{\rm keV}^{-1}]
\end{equation}

The best fit, shown in Fig.~\ref{fig:xall}, yields values
 (and 1\,$\sigma$ errors) of: $C= (10.15\pm0.80)\times 10^{-2} $, 
$\Gamma_1=1.32\pm0.018$, $\Gamma_2=2.88\pm0.015$
and $E_{\rm B}=29.99\pm1.1$\,keV. 
The reduced $\chi^2$ is acceptable ($\sim$1.2)
considering the number (10) of different datasets fitted. The suggested 
formula reproduces well the CXB spectrum over two decades in flux 
and five in energy. At a given energy, the systematic uncertainty
produced by the scatter of the measurements used here is of the 
of the order of 10\,\%.

Note that, there is no astrophysical need 
to connect  the keV and the GeV 
diffuse backgrounds with a single formula \citep[e.g.][]{gruber99}.
It is generally agreed that
the source populations contributing to the two diffuse
backgrounds are probably different. Almost all of the 
CXB radiation up to 300\,keV can be explained in terms of 
emission-line AGN \citep[e.g.][]{gilli07}. Moreover, 
taking into account (the likely, but not yet detected) population
of non-thermal electrons in AGN coronae, \cite{inoue08}
successfully reproduce the CXB emission up to 4\,MeV.
On the other hand, blazars account only for $\leq25$\,\%
of the GeV diffuse background and most likely other source
classes contribute to the diffuse emission \citep{dermer07}.

%%%%%%%%%%%%%%%%%%%%%%%%%%%%%% --------- Fig 14

%%%%%%%%%%%%%%%%%%%%%%%%%%%%%%%%%%%%%%%%%%%%%%%%%%%%%%%%%%%%%%%%%%%%%%%%%%%%%
%%%%%%%%%%%%%%%%%%%%%%%%%%%%%%%%%%%%%%%%%%%%%%%%%%%%%%%%%%%%%%%%%%%%%%%%%%%%%
\subsection{The Earth albedo Spectrum}\label{sec:albedospec}

Our Earth albedo spectrum  is not compatible with the classical high-energy
photon index of 1.4. The BAT data are
consistent with a steeper high-energy photon index at 
99.989 confidence level.

\begin{figure}[ht!]
  \begin{center}
  	 \includegraphics[scale=0.89,angle=0.]{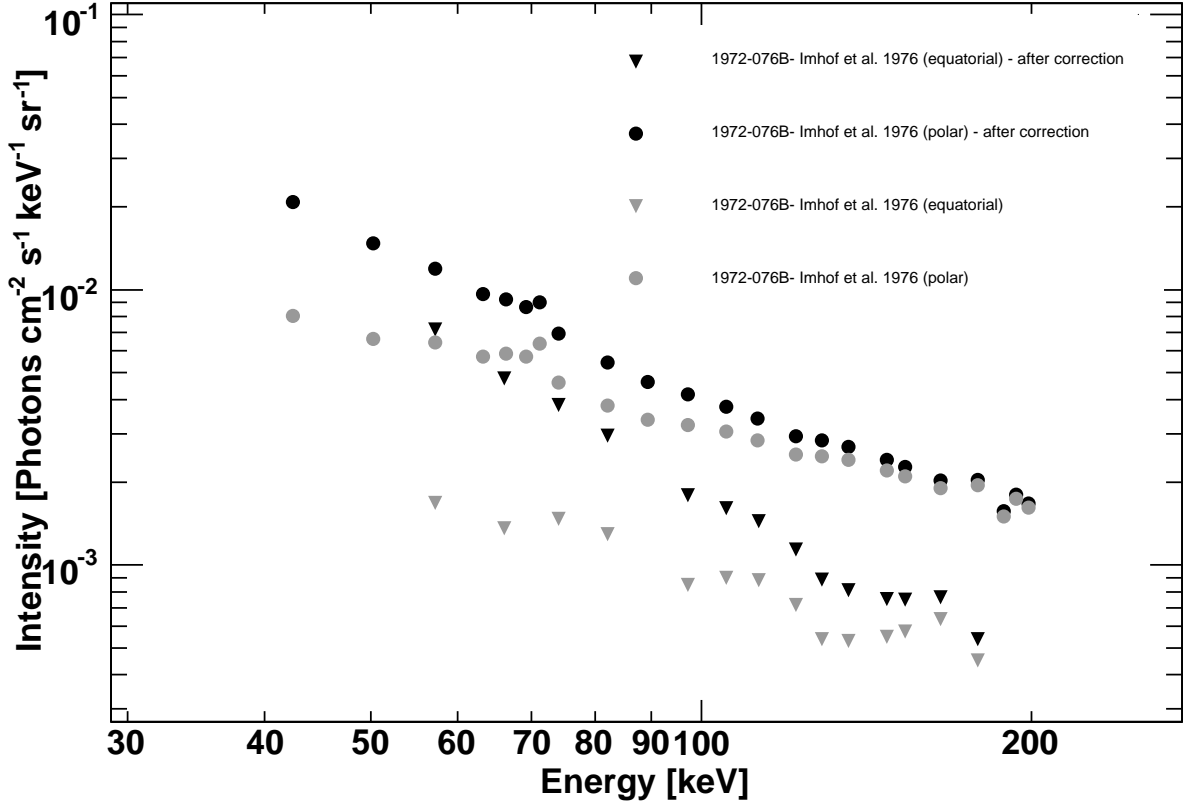}
  \end{center}
  \caption{Measurements of the Earth albedo of \cite{imhof76}
at different geomagnetic latitudes. The original measurements
(gray) have been corrected for un-subtracted CXB emission (see the text
for details). After the correction, the albedo spectra (black data points)
become steeper. 
 \label{fig:imhof}}
\end{figure}

Using the   polar-orbiting satellite {\it 1972-076B},
\cite{imhof76} found that above 40\,keV the photon spectrum is 
consistent with a power-law with an index ranging from 1.34 to 1.4,
depending on the latitude range scanned. Their measurement is based
on the difference between pointings towards the atmosphere ({\it Down}) and 
pointing towards the sky ({\it Up}). In order to derive the albedo spectrum,
the authors sum the CXB emission and  the {\it Down-Up} spectrum 
\citep[see Eq.~5 in][for details]{imhof76}.
For the CXB emission, they adopt the measurement of \cite{pal73},
which describes the CXB photon spectrum as $dN/dE = 25E^{-2.1}$.
This representation differs from the HEAO-1 CXB spectrum
in both normalization and photon index
in the 40--200\,keV range.
Thus, we adjusted the \cite{imhof76} albedo spectra, 
taking into account the 
differences between the \cite{gruber99} and the \cite{pal73} CXB 
spectral representations.
This is shown in  Fig.~\ref{fig:imhof}.
After the correction, the two (equatorial and polar) Albedo spectra
are consistent with a power-law with  photon index $\sim$1.7.
In particular the equatorial  measurement is in good agreement
with the BAT spectrum.

%%%%%%%%%%%%%%%%%%%%%%%%%%%%%% --------- Fig 15

The  {\it BeppoSAX} satellite operated in a Low-Earth Orbit
similar to  {\it Swift}, but with different 
inclination (4$^{\circ}$).  The BAT Earth albedo spectrum is compatible 
(within the large uncertainties of the {\it BeppoSAX} analysis)
with the measurements obtained by \cite{frontera07}.
It is worth noting that  we derived
the Earth intensity using different orbital positions
(as done for {\it BeppoSAX} and {\it OSO-3}), and thus averaging
over the magnetic latitude sampled by {\em Swift}. 
Moreover, as generally the Earth enters the FOV
at large angles, we do not observe the upward albedo, but rather the albedo 
emerging at large zenith angles. This also seems confirmed 
by the similarity of our spectrum with the {\it downward } gamma-ray
flux measured for a balloon over Palestine, Texas \citep{gehrels85}.

\begin{figure}[ht!]
\begin{center}
\includegraphics[scale=0.90,angle=0]{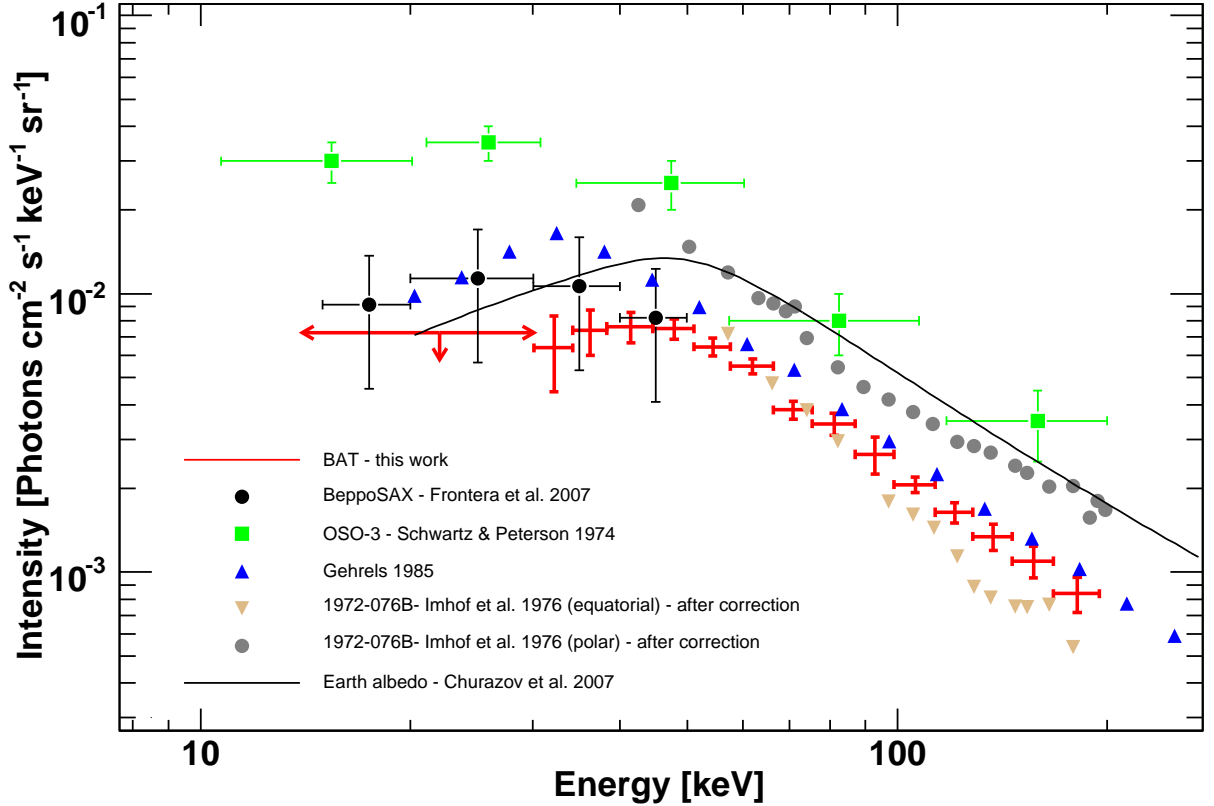}
\end{center}
\caption{BAT Earth spectrum  as compared
to past measurements. References are shown in the legend.
The datapoints (triangles) from \cite{gehrels85} are a fit to the
{\it downward} gamma-ray flux at 5\,g cm$^{-2}$ over Palestine, Texas.
The measurements from \cite{imhof76} were corrected to take into account
the correct CXB emission (details are in the text).
The thin solid line shows the prediction of the Earth emission as
observed from the orbit of the {\it INTEGRAL} satellite \citep{sazonov07b,churazov07}.
 \label{fig:ea_spec}}
\end{figure}

Fig.~\ref{fig:ea_spec} reports also the prediction of the Earth albedo
emission as observed from the orbit of the {\it INTEGRAL} satellite
\citep{sazonov07b}.
Its normalization has been derived during the measurement
of the CXB intensity \citep{churazov07}. It is evident
that this prediction and the BAT measurement agree well in shape but not
so in normalization. Among many factors, the overall normalization depends 
strongly on the geomagnetic latitude and the distance to the Earth.
The agreement of the BAT and {\it INTEGRAL} albedo spectra, respectively, with the
equatorial and polar measurements of \cite{imhof76} seem to confirm this
interpretation.
%%%%%%%%%%%%%%%%%%%%%%%%%%%%%% --------- Fig 16

%%%%%%%%%%%%%%%%%%%%%%%%%%%%%%%%%%%%%%%%%%%%%%%%%%%%%%%%%%%%%%%%%%%%%%%%%%%%%
%%%%%%%%%%%%%%%%%%%%%%%%%%%%%%%%%%%%%%%%%%%%%%%%%%%%%%%%%%%%%%%%%%%%%%%%%%%%%
\section{Discussion}
We have used Earth occultation episodes to derive with {\it Swift}/BAT 
an accurate measurement of the CXB emission in the 15--200\,keV energy
range. Moreover, we have proven by means of an independent technique 
the accuracy of the occultation analysis and of our results.
The observed BAT intensity near the peak of the CXB spectrum at 30\,keV
is 46.2\,keV$^2$ cm$^{-2}$ s$^{-1}$ sr$^{-1}$ keV$^{-1}$ and 
its uncertainty\footnotemark{}
\footnotetext{Derived combining both measurements of the CXB.}
is $\sim$3\,\% (including all systematics). The normalization of the BAT
CXB spectrum at 30\,keV is $\sim8$\,\% larger than the HEAO-1 \citep{gruber99}
measurement  and consistent  with the {\it INTEGRAL} one \citep{churazov07}.
Moreover, considering that the precision of the HEAO-1 measurement at the
CXB peak is 10\,\% \citep{marshall80} and that {\it BeppoSAX} data
are compatible with a larger (up to 20\,\%) normalization of the CXB
spectrum shows that all measurements above 10\,keV are consistent
within their systematic uncertainties.
%%%%%

Such consistency is not observed 
at lower energies \citep[e.g. see discussion in ][]{revnivtsev05}. 
The origin of this inconsistency
is unclear. However, it seems that neither cosmic variance \citep{barcons00}
nor differences in the flux scale calibration of each individual 
instrument  \citep{revnivtsev05,frontera07} may account for it.
A likely reason for the discrepancy of CXB measurement in the 2--10\,keV
band might reside in 
a systematic error in the response function used for diffuse
sources  \citep{frontera07}. 
To our knowledge, BAT is the only
instrument (beside HEAO-1 A2 which was designed with the purpose
of measuring the CXB) which makes use of a dedicated instrumental
response developed for this particular analysis.
We also note that a recent measurement of the CXB performed, in the 2--7\,keV,
by {\it Swift}/XRT \citep{moretti08} seems to confirm
the results  of  XMM-Newton \citep{deluca04}, RXTE \citep{revnivtsev05}
and Chandra \citep{hickox06}. If confirmed, this  means
that the CXB spectrum, as most recently measured, 
is 25--40\,\% larger (with respect the measurement of 
HEAO-1) below 10\,keV while only $\sim$10\,\%  larger above 20\,keV.
The functional form we provide in $\S$~\ref{sec:cxbspec} for the broad-band
CXB emission approximates well this scenario. 
%%%%
%%%%

A larger, than previously estimated, CXB emission would in turn
require a larger density of Compton-thick AGN both in the local and in the 
more distant Universe. 
Recently, {\em Chandra} stacking analyses of 
mid-IR selected sources  unveiled  a large
population of Compton-thick AGN at high redshift \cite[][]{daddi07,fiore08}.
This large fraction of Compton-thick AGN found at
z=$\sim2$ can be accommodated if the emitted (obscured) flux
is very low. Indeed, lowering the assumed scattering efficiency 
(ratio of reflected to nuclear flux) 
would increase by the same amount the number of Compton-thick AGN at 
any redshift. In this framework, the recent discoveries of Compton-thick
AGN with an extremely low scattering efficiency 
\citep[][]{ueda07,comastri07} fits well. These
AGN are likely buried in a geometrically thick torus that obscures most
of the nuclear flux. Although there can be many of these hidden AGN,
their individual contribution to the CXB is necessarily small.
This seems to be confirmed by the fact that the
contribution of the mid-IR selected, z=$\sim2$, AGN 
is $<3$\% of the CXB intensity in the 10--30\,keV band \citep{daddi07}. 
Therefore,
a larger contribution should be provided by Compton-thick AGN 
at lower redshift.

% 
%%%%%%%%%%%%%%%%%%%%%%%%%%%%%%%%%%%%%%%%%%%%%%%%%%%%%%%%%%%%%%%%%%%%%%%%%%%%%
%
%
\section{Conclusions}
BAT performed a very sensitive measurement of the CXB emission
in the 15--200\,keV energy range.
This measurement takes advantage of several episodes of CXB flux
modulation due to Earth's passages through the BAT FOV.
We find that the BAT CXB spectrum is in good agreement with the {\it INTEGRAL}
one and that its normalization is  $\sim8$\,\% larger than the HEAO-1
measurement at 30\,keV. 
Additionally, performing an independent measurement of the CXB in the 
15--50\,keV band, we are able to confirm this result.
Remarkably, our
 study  also shows that all the available measurements in the $>10$\,keV
range agree within their systematic uncertainties.
The new analyses of {\it COMPTEL} and  {\it EGRET} data 
\citep{weidenspointner00,strong04} show that the formula
suggested by \cite{gruber99} for the diffuse X- and gamma-ray backgrounds
is only valid below 2\,MeV. We derived a simple functional form which, in
the 2--2000\,keV range, approximate well (to a precision of 10\,\%)
the CXB spectrum. 

Our study also derives the Earth albedo spectrum averaged over the magnetic
latitudes sampled by {\em Swift}. 
The BAT spectrum is in agreement with all the  previous observations performed
by satellites operating in similar LEO orbits.
This work shows that the Earth albedo spectrum declines at energies $>$40\,keV
according to 
 a power-law with photon index of $\sim$1.7, and not as 1.4 as previously
thought. A re-analyis of the measurements performed by \cite{imhof76} 
is in perfect agreement with the BAT Earth albedo spectrum.
The good agreement among  the available measurements 
allows to use the BAT Earth albedo spectrum
to predict the background contribution from the Earth for other instruments
operating at similar orbits.

%%%%%%%%%%%%%%%%%%%%%%%%%%%%%%%%%%%%%%%%%%%%%%%%%%%%%%%%%%%%%%%%%%%%%%%%%%%%%%
\acknowledgments
We are grateful to S. Barthelmy, J. Cummings and H. Krimm for 
all the effort spent in keeping the BAT perfectly operating.
MA acknowledges  the useful suggestions of  R. Mushotzky and 
the help of C. Gordon for adapting Xspec to the purposes
of this analysis. The anonymous referee is aknowledged for 
his/her helpful comments which improved the manuscript.
MA is grateful to  S. Sazonov and A. Zoglauer for interesting
discussions about the Earth emission.
This research has made use of data obtained from the 
High Energy Astrophysics Science Archive Research Center (HEASARC) provided 
by NASA's Goddard Space Flight Center.
MA acknowledges funding from the DFG Leibniz-Prize to GH (HA 1850/28-1).

\bibliographystyle{apj}
\bibliography{/Users/marcoajello/Work/Papers/BiblioLib/biblio}

\end{document}